\documentclass{aa}
\pdfoutput=1
\usepackage{graphicx}   
\usepackage{subfigure}
\usepackage[varg]{txfonts}
\bibpunct{(}{)}{;}{a}{}{, } 
\usepackage{longtable}

\newcommand{\Ma}{\ensuremath{M_\mathrm{a}}}
\newcommand{\Mb}{\ensuremath{M_\mathrm{b}}}

\newcommand{\Ka}{\ensuremath{K_\mathrm{a}}}
\newcommand{\Kb}{\ensuremath{K_\mathrm{b}}}
\newcommand{\Aap}{\ensuremath{a_\mathrm{app}}}

\newcommand{\Msun}{\ensuremath{\mathrm{M_{\sun}}}}

\newcommand{\mas}{\ensuremath{\mathrm{\, mas}}}
\usepackage{color}

\begin{document}

\title{The VLTI / PIONIER  near-infrared interferometric survey of southern T~Tauri stars. I. First results\thanks{Data obtained at the ESO VLTI as part of programmes  086. C-0433, 087. C-0703, 088. C-0670, and 089. C-0769. }.
 }


\author{F. Anthonioz\inst{1}
\and F. M\'enard\inst{2}\inst{, 1}
\and C. Pinte \inst{1}
\and J-B. Le Bouquin \inst{1}
\and M. Benisty \inst{1}
\and W. -F. Thi \inst{1}
\and O. Absil \inst{4}
\and G. Duch\^ene\inst{5}\inst{, 1}
\and J.-C. Augereau \inst{1}
\and J. -P. Berger\inst{3}
\and S. Casassus \inst{8}
\and G. Duvert \inst{1}
\and B. Lazareff \inst{1}
\and F. Malbet\inst{1}
\and R. Millan-Gabet\inst{6}
\and M.R. Schreiber\inst{9} 
\and W. Traub\inst{6}\inst{, 7}
\and G. Zins \inst{1}
}

\institute{UJF-Grenoble 1 / CNRS-INSU, Institut de Planetologie et d’Astrophysique de Grenoble (IPAG) UMR 5274, Grenoble, F-38041, France
\and UMI-FCA, CNRS/INSU France (UMI 3386) , and Universidad de Chile, Santiago, Chile
\and European Southern Observatory, D-85748, Garching by M\"unchen, Germany
\and D\'epartement d'Astrophysique, G\'eophysique et Oc\'eanographie, Universit\'e de Li\`ege, 17 All\'ee du Six Ao\^ut, B-4000 Li\`ege, Belgium
\and Astronomy Department, University of California, Berkeley, CA 94720-3411 USA
\and California Institute of Technology, Pasadena, CA 91125, USA. 
\and Jet Propulsion Laboratory, California Institute of Technology,Pasadena, CA, 91109, USA. 
\and Departamento de Astronom\'ia, Universidad de Chile, Casilla 36-D, Santiago, Chile.
\and Departamento de Fisica y Astronom\'ia, Universidad de Valpara\'iso, Valpara\'iso, Chile 
}

\date{accepted December 2014}

\abstract {The properties of the inner disks of bright Herbig AeBe stars have been studied with near infrared (NIR) interferometry and high resolution spectroscopy. The continuum (dust) and a few molecular gas species have been studied close to the central star; however,  sensitivity problems limit direct information about the inner disks of the fainter T Tauri stars.}
{ Our aim is to measure some of the properties (inner radius, brightness profile, shape) of the inner regions of circumstellar disk surrounding southern T~Tauri stars. }
{ We performed a survey with the VLTI/PIONIER recombiner instrument at H-band of 21 T~Tauri stars. The baselines used ranged from 11 m to 129 m, corresponding to a maximum resolution of $\sim$3mas ($\sim$ 0.45 au at 150 pc). } 
{Thirteen disks are resolved well and the visibility curves are fully
sampled as a function of baseline in the range 45-130 m for these 13 objects. 
A simple qualitative examination of visibility profiles allows us to identify a rapid drop-off in the visibilities at short baselines(<10M$\lambda$) in 8 resolved disks. This is indicative of a significant contribution from an extended (R > 3 au, at 150 pc) contribution of light from the disk. We demonstrate that this component is compatible with scattered light, providing strong support to a prediction made by Pinte et al. (2008). 
The amplitude of the drop-off and the amount of dust thermal emission changes from source to source suggesting that each disk is different. A by-product of the survey is the identification of a new milli-arcsec separation binary: WW Cha. Spectroscopic and interferometric data of AK Sco have also been fitted with a binary + disk model.  }
{The visibility data are reproduced well when thermal emission and scattering from dust are fully considered. The inner radii measured are consistent with the expected dust sublimation radii. The modelling of AK Sco suggests a likely coplanarity between the disk and the binary's orbital plane. }

\maketitle

\section{Introduction}
Gas-rich circumstellar disks around young stars (also known as protoplanetary disks) are central to the formation process of both stars and planets.  
They contain the mass reservoir to fuel accretion onto the central star, they are the vector by which angular momentum is evacuated by bipolar outflows, and they are the sites where planetesimals grow and planets form. Based on direct imaging of planetary-mass bodies embedded in debris disks ($\beta$ Pic, \cite{Lagrange2010}; Fomalhaut, \cite{Kalas2008};  HR 8799, \cite{Marois10}) and on the coplanarity in the solar system and extrasolar systems \citep{Figueira12}, 
there is now little doubt that planets do form in disks.

The inner central regions (R < 10 au) of these protoplanetary disks are difficult to observe directly because of their small apparent size. Unfortunately, this is where the density is high enough for rocky terrestrial planets and gas-giant embryos to form within reasonable timescales compared to the disk lifetime. A knowledge of the geometry, temperature, and content of these regions is critical for understanding how mass is transferred onto the star and how planets may form, agglomerate, and migrate. 

Near-infrared long-baseline interferometry has, in principle, the necessary angular resolution to resolve these regions, those located in the range 0.5-10 au from the central star at the distance of the nearest star forming regions (i.e., d=140 pc). However, to obtain reliable or detailed information on the disk location and shape, a good two-dimensional spatial frequency (hereafter  uv) coverage is needed. This is not easily available, in particular for faint targets. 

Because they are relatively bright, Herbig AeBe stars have been amply observed by near infrared (NIR) interferometers in the past, and the dust and gas distributions in the inner regions of their disk are now reasonably well characterised  (see, e.g.,  \cite{Dullemond10} and references therein for an exhaustive review of the inner disks around Herbig stars). Interestingly, for these stars, a fairly tight correlation is found between the luminosity of the central star and the characteristic radius that the disk emission comes from \citep{Monnier02}. This radius was rapidly associated to the dust sublimation radius.

On the other hand, observations are much less common for the fainter solar-like counterparts of Herbig stars, the T Tauri stars, because interferometers are usually not sensitive enough or have a limited number of baselines available. Also, because of the lower luminosity and temperature of the central T Tauri stars, the inner rim of their dust disks, typically 0.1 au in radius corresponding to $\sim$1 mas at 150 pc,  is located closer to the centre compared to Herbig stars. This is more challenging to resolve and, as a consequence, the inner dust and gas distributions are less well constrained than for the bright Herbig stars. 

While current NIR interferometers, with their 100--200 m baselines, do not have the necessary resolution to fully resolve the inner disks of T Tauri's, such as measuring the first zero of their visibility curves, they can still provide useful data for constraining the inner disk radius. Previous studies have estimated the inner rim radii of T Tauri disks using ring-like disk models (\cite{Eisner07,Eisner14} and references  therein). These were  appropriate given the available data. Interestingly, these estimations of radii depart significantly from the correlation found for the Herbig stars between the inner rim size and the luminosity \citep{Eisner07}. Several explanations were put forward to explain this departure: for example,  peculiar dust properties, large magnetospheric radii \citep{Eisner07}. \cite{pinte08} suggest, however, that this departure could also be explained by taking scattered light into account in the modelling effort, i.e., using full radiative transfer including light scattering and thermal emission rather than dust thermal re-emission alone. They show for that case from a generic model, that T~Tauri stars can be p put back on the expected correlation between luminosity and disk sublimation radius.  One goal of this paper is to verify that suggestion further. Detailed fitting of individual targets was not done, however, and these predictions could not be verified owing to the limited sampling of the uv plane. This limited coverage results in significant ambiguities in the models. These ambiguities can be mitigated or solved by a broader coverage in baseline lengths and orientation.

PIONIER \citep{Lebouquin11} at the ESO-VLTI offers the possibility to recombine the light from four telescopes at once. This recombiner is also more sensitive than previous ones, allowing good measurements to be obtained for the fainter T~Tauri stars. Interferometric observations with four telescopes provide six independent baseline measurements at once, as well as three independent closure phases. This is an improvement over previous interferometric recombiners in terms of sensitivity and rapid uv coverage. 

In this paper we report observations of 21 T~Tauri stars and 2 Herbig stars from the southern hemisphere with the interferometric instrument PIONIER. In section~\ref{sec : obs}, we give more details about the observations. Sections ~\ref{sec : nondetec},~\ref{sec:viscurv}, and \ref{sec : mod data} are devoted to the statistical results for non-detection and generic modelling of the visibilities of our sample. Section~\ref{sec : bin} is devoted to  binaries, and we conclude in section~\ref{sec : fin}. 

\section{Observations, data reduction, and sample}
\label{sec : obs}
\subsection{The sample}

The sample comprises 21 T~Tauri stars (spectral type G or later) and 2 Herbig Ae stars (spectral types F and A) brighter than H = 8.5. This is the current limiting magnitude of PIONIER for good seeing conditions, which is H$\sim$8.0 for median conditions. In addition to the limiting magnitude, the criteria used for selection were 
\begin{itemize}
\item a significant NIR excess that traces hot dust located close to the star, 
\item  a resolved image from radio interferometry tracing the colder dust located in the outer disk, or
\item a scattered light image. 
\end{itemize}

All the targets are located in southern star forming fegions and young associations: six are located in the Lupus associations, four in the TW Hya Association (or co-moving group), three in the $\rho$ Oph cloud, three  in Upper Scorpius, two in CrA, and one each in the Upper Centaurus Lupus, Argus, Chamaleon I, and $\beta$ Pic moving groups\footnote{We take the result with the higher membership probability from \cite{Malo13} for FK Ser and V4046 Sgr.}. 
The remaining star is located in Orion.
The coordinates, spectral type, distance, and magnitude in H for these stars are summarised in Table~\ref{tab : dist_star}. Five of these stars (AS 205 A, V2129 Oph, V2508 Oph, S CrA, and TW Hya) have been previously observed by NIR interferometry \citep{Eisner05, Eisner07,Eisner10,Vural12,Menu14}

\subsection{Known and new binaries}

In our sample, 12 objects are previously known binaries or multiple systems. For 6 of them the companion lies at a separation large enough  to not be included in the field of view of PIONIER, which is approximated by the FWHM of the fibre response function, i.e., 250mas. These can be considered as made of two separate single stars. We observed the two members of the TWA~3 system separately. 

For the six other targets, the companion is included in the field of view. The companion of V4046 Sgr has a separation of 0.56 mas, which is much less than the resolving power of PIONIER. It is not resolved by our observations. The same applies for V380~Ori C. HT~Lup has two companions at 2.8" and 0.126".  HT Lup C is possibly detected in our short baseline observations, while there is no indication of its presence on the long baseline observations. This object will be discussed more in detail in a dedicated paper. Finally, V380 Ori B and HN Lup B are not detected in our observations because they would produce large closure phases up to 60\degr and $\sim$30\degr , respectively, while the observed closure phases are compatible with 0.   A summary of the flux ratio and separation of the multiple systems studied here is given in Table~\ref{tab : binaires}. Our survey also reveals the binarity of WW Cha. For this star and the three last known binaries (AK Sco, V1000 Sco, and TWA 3A), the companion is close enough to disrupt the inner part of the disk, meaning that these objects do not follow any (simple or not) size-luminosity relationship.  We thus discuss these four targets separately from the rest of the sample, in \S~\ref{sec : bin}.\\

\begin{table*}[t]
\caption{\label{tab : dist_star} Position,  spectral type, distance, H magnitude, and binarity of the sample.  For the binarity, "\textbf{Unresolved}" means that the companion is unresolved by PIONIER,"\textbf{Yes}" means that the star has a detected companion in PIONIER's  half field of view (125 mas), "\textbf{Border}" that the companion lies on the edge of the field of view, stars with "\textbf{Outside}" have a companion with a separation much larger than the field of view, and "\textbf{no}" indicates that the star is single.
}
\centering
\begin{tabular}{llllrrrrrr}
\hline
\hline
Star&R. A. &Dec&SpT& dist.(pc)&Log($L/L_{\sun}$)&Refs&H-mag&Resolved?&binary?\\

\noalign{\smallskip}\hline\noalign{\smallskip}
V380 Ori        &05 36 25       &       -06 42 57       &A1e    &510            &1.99 &       1,2,3   &6.96&  yes&            Unresolved,Border\\
TWA 07          &10 42 30       &       -33 40 16       &M3.2   &34                     &-0.94& 4       &7.13&   no&             no\\
TW Hya          &11 01 51       &       -34 42 17       &M0.5   &56                     &-0.72& 4       &7.55&  marginally&     no\\
WW Cha          &11 10 00       &       -76 34 57       &K5             &160            &0.74&  5,6     &7.21&  yes&            Yes\\
TWA 3A          &11 10 28       &       -37 31 52       &M4.1   &35             &-0.92& 4       &7.53&  yes&            Yes, Outside\\
TWA 3B          &11 10 28       &       -37 31 52       &M4.0   &35             &-1.10& 4       &8.15&  no&                     Outside\\
HT Lup          &15 45 12       &       -34 17 30       &K2             &150            &0.74&  4       &6.87&  yes&            Border, Outside\\               
HN Lup          &15 48 05       &       -35 15 52       &M1.5   &150            &-0.28& 7,14    &8.1 &       yes&            Outside\\
GQ Lup          &15 49 12       &       -35 39 05       &K7             &150            &0.17&  27      &7.70&  yes&            Outside\\
RU Lup          &15 56 42       &       -37 49 15       &K7             &150            &0.16&  8,9,22  &7.82&  yes&            no\\
V1149 Sco       &15 58 36       &       -22 57 15       &G6             &145            &0.39&  10      &7.69&  yes&            no\\
RY Lup          &15 59 28       &       -40 21 51       &G0V    &150            &0.41    &       2,11,12 &7.69&  yes&            no\\
MY Lup          &16 00 44       &       -41 55 31       &K0             &150            &-0.20& 13,23   &8.69&   no&             no\\
V1000 Sco       &16 11 08       &       -19 04 46       &K2             &145            &0.44&  14,24   &7.98&  yes&            Yes\\
AS 205 A                &16 11 31       &       -18 38 24       &K5             &125            &0.60&  15      &6.75&  yes&            Outside\\
V2129 Oph       &16 27 40       &       -24 22 04       &K5             &121            &0.15&  26      &7.67&  yes&            no\\
V2508 Oph       &16 48 45       &       -14 16 35       &K6             &125            &0.46&  16,17   &7.57&  yes&            no\\
V1121 Oph       &16 49 15       &       -14 22 08       &K5             &130            &0.176& 2,12    &7.45&  yes&            no\\
AK Sco          &16 54 44       &       -36 53 18       &F5V    &145            &0.61$\times$2& 18,22   &7.06&  yes&            Yes\\
V4046 Sgr       &18 14 10       &       -32 47 34       &K5V    &73                     &-0.41$\times$2&        19      &7.44&  no&                     Unresolved\\
FK Ser          &18 20 22       &       -10 11 13       &K6IV   &32                     &0.2&   2,20,25 &6.92&  no&                     Outside\\
S CrA N         &19 01 08       &       -36 57 19       &K3     &130            &0.36&  15      &7.05&  yes&            Outside\\
V709 CrA        &19 01 34       &       -37 00 56       &K1IV   &130            &0.19&  21      &7.97&  no&                     no\\
\hline  
\end{tabular}
\tablebib{ (1)~\cite{Manoj06}, (2)~\cite{VanLeeuwen07}, (3)~\cite{Alecian13}, (4)\cite{Herczeg14}, (5)\cite{Luhman07}, (6)~\cite{Whittet97},(7)\cite{hughes94}, (8)\cite{Lommen07}, (9)\cite{Stempels07}, (10)~\cite{Yang12}, (11)~\cite{Reipurth96}, (12)~\cite{Artemenko12}, (13) \cite{Romero12}, (14)~\cite{Sartori03}, (15)\cite{Bast11}, (16)\cite{Andrews10}, (17) \cite{DeGeus89}, (18)~\cite{alencar03}, (19)~\cite{Donati11},(20)~\cite{Torres06}, (21)~\cite{Forbrich07}, (22) this paper, (23)~\cite{Gregorio-Hetem02}, (24)~\cite{Wahhaj10}, (25)~\cite{Mcdonald12}, (26)~\cite{Donati11}, (27)~\cite{Dai10}}
\end{table*}

\begin{table}[ht]
\caption{\label{tab : binaires} Separation, luminosity ratio and references for the observed binaries.
}
\centering
\begin{tabular}{llcl}
\hline
\hline
Star&Sep. (mas) & L $_{ (comp) } $/L $_{ (prim) } $& Refs. \\

\noalign{\smallskip}\hline\noalign{\smallskip}
&&\\
WW Cha          &       6.31 $\pm $0.16         &       0.6                                                     & (1) \\
V 1000 Sco      &        $4.14\pm $0.25         &       0.47                                            & (1),(12) \\
TWA 3A          &       3.51 $\pm $0.5  &       0.76                                            & (1) \\
AK Sco          &       1.11 $\pm$0.04          &       $\sim $1                                        & (1),(2) \\
V4046 Sgr       &       0.56                            &       0.67                                            & (9) \\
HT Lup          &       126 $\pm $1                     &       0.15                                            & (3) \\
V380 Ori        &       154 $\pm $2                     &       0.31 $\pm $0.01                           & (4) \\
                        &       <0.33 $\pm $2           &       0.031 $^{+0.16}_{-0.026} $       & (10) \\

\noalign{\smallskip}\hline\noalign{\smallskip}

GQ Lup          &       732                                     &       $\sim $0.004                          & (5) \\
HN Lup          &       240 $\pm $10            &       0.4 $\pm $0.02                          & (3) \\
HT Lup& 2800 $\pm $100          &       0.095 $\pm $0.005                                       & (3) \\
S CrA           &       14100 $\pm $60          &       0.3 $\pm $0.02                          & (3) \\
AS 205          &       1400                            &       0.31 $\pm $0.01                           & (6) \\
TWA 3A-B                &       1440 $\pm $10           &       $\sim $0.63                             & (8) \\
FK Ser          &       1330                            &                                                               & (11) \\

\hline

&&\\
\end{tabular}

\tablebib{ (1)~this paper; (2)~\cite{andersen89}; (3)~\cite{ghez97}; (4) \cite{leinert94}; ; (5)~\cite{mugraueur05}; (6)~\cite{cohen79}; (7)~\cite{dyck82}; (8)~\cite{delareza89}; (9)~\cite{byrne86}; (10)~\cite{Alecian09}; (11)~\cite{Herbig73}; (12)~\cite{Mathieu89} }
\end{table}

\subsection{The observations}

The observations were performed with the PIONIER 4-telescope beam recombiner instrument \citep{Lebouquin11} 
using the four 1.8 m Auxiliary Telescopes (AT) of the Very Large Telescope Interferometer (VLTI, \cite{Haguenauer10})
at the Paranal Observatory of the European Southern Observatory (ESO) during five different semesters from period P86 to P90. The observations were obtained in visitor mode. In total, 17.5 
nights were allocated to the programme with a long-baseline configuration distributed in nine sub-runs over four semesters (P86-P89). Seven more nights were allocated with short a baseline configuration during P90. For the long baseline survey, seven full nights (40\%) were lost from adverse weather conditions, the weather conditions being average for the remaining 10.5 nights. For the short baseline run, nearly all the observable time (6.5 nights out of 7) was lost due to weather, and only 4 stars could be observed.  For Period 86, the stations were A0-K0-G1-I1
 and for P87, P88, and P89 the stations were A1-K0-G1-I1, providing separations on the ground between telescopes ranging from 47 meters to 129 metres, equivalent to a maximum angular resolution of $\sim3 $~milliarcseconds. For period P90, the stations were  A1-B2-C1-D0
 A log of the observations is presented in Appendix \ref{app : log}. 

The observation strategy was designed to interleave the science target between different interferometric calibrators as much as possible. A typical observation sequence (5 blocks) was Calibrator 1 --- Science Target --- Calibrator 2 --- Science Target  ---  Calibrator 1. The calibrators were chosen from the JSDC \citep{Lafrasse10} and selected to be unresolved single stars. The calibrators have H-magnitudes that are usually a little brighter than the science targets (between 0.0 and 0.75 H-mag brighter). 
 Each block, either science or calibrator, was composed of 5 or 10 exposures, each of which composed of 100 fringe scans, followed by the acquisition of the dark frame  and the internal, flat-fielded flux splitting ratio \citep{Lebouquin11}. 
Data were reduced and calibrated with the dedicated {\sc pndrs} package \citep{Lebouquin11}. 
Typical errors on our measurements are $\sim $5\% for the visibilities  and $\sim $3.5$\degr$ on the closure phases. The final, calibrated interferometric data acquired during this survey are presented in Appendix \ref{app:images}. Our observations can be roughly separated into three distinct groups, depending on the main signature of interferometric data:  unresolved targets, binaries, and stars with a resolved disk. 

\section{Unresolved targets}
\label{sec : nondetec}
\begin{center}
\begin{figure*}[ht]
\includegraphics[width=15cm]{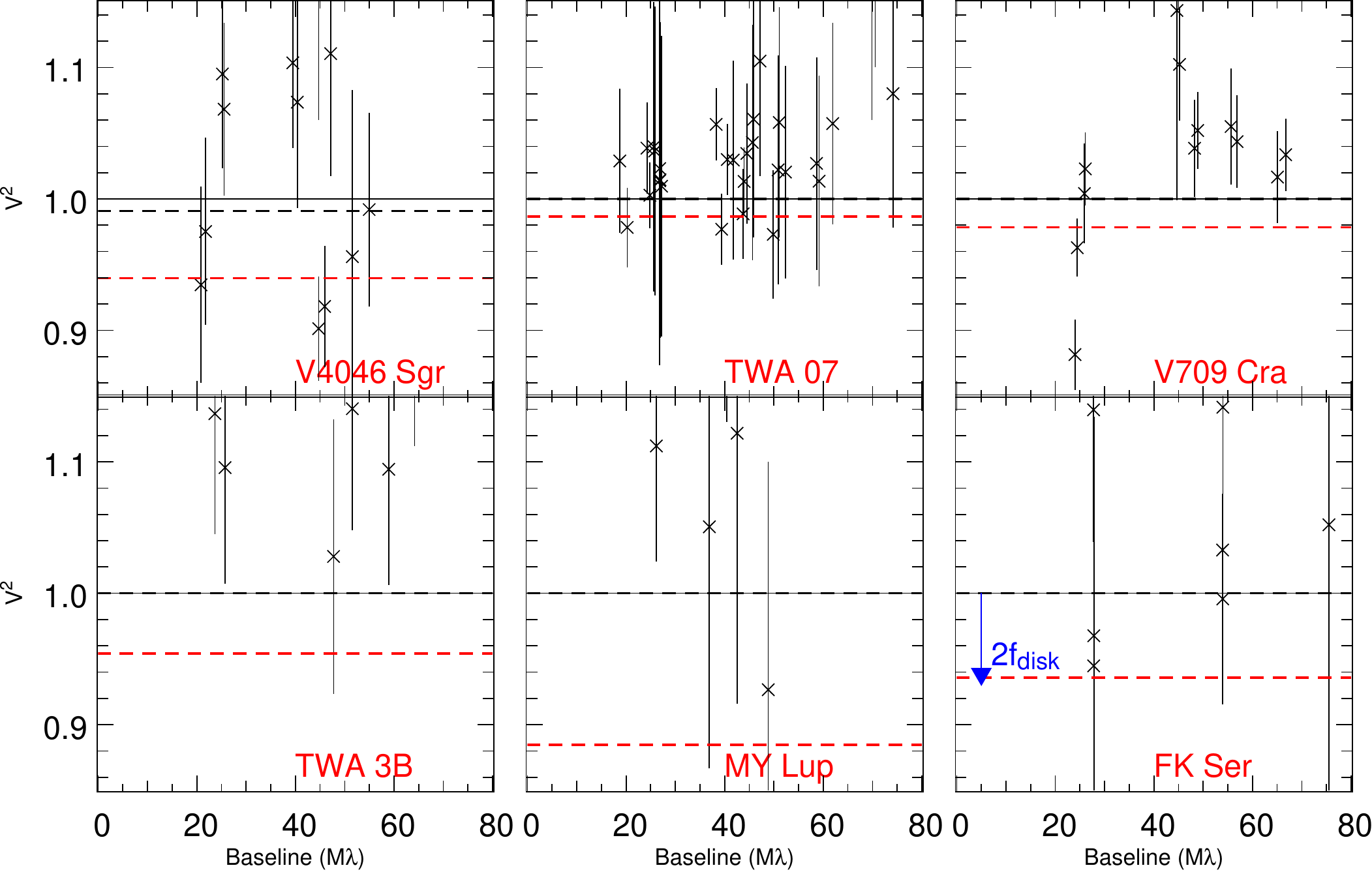}
\caption{\label{fig : fit nondetec} Visibility profile modelling of the unresolved stars of our survey. The visibility data correspond to the black crosses and error bars, while the best visibility fit and lower limit are plotted in dashed black and red lines respectively. The difference between unity and the lower limit can be approximated to two time the maximal fractional luminosity of the disk.}
\end{figure*}
\end{center}

Six targets in the survey are unresolved, meaning that their squared visibility at the longest baselines are compatible with unity. To set limits on the maximum brightness of their disks, we follow a similar linear fitting procedure, as described in \cite{Difolco07}. The procedure consists of fitting the visibility data with a model made of a central star surrounded by a faint, fully resolved uniform disk. Diameters of the stars are expected to be 0.22 mas or less, so are unresolved even at the longest baselines. The visibility, V, as a function of baseline, B, can therefore be written as 

\begin{equation}
V^2 (B) =\left(\dfrac{V_{\star}}{1+\mathrm{f_{disk}}}  \right)^2 \approx 1-2\mathrm{f_{disk}}
\label{eq : Fmax}
,\end{equation}
where $f_{disk}= F_{disk}/F_{tot}$. We then calculate the associated probability of each model
\begin{equation}
 p(f_{disk})  \propto  e^{-\chi^2(f_{disk})/2}  
\label{eq : proba_Fmax}
\end{equation}
and define the confidence interval as the range over which the cumulative probability is greater than 99.6\%.
The results of this visibility fitting are displayed Figure~\ref{fig : fit nondetec}. The maximum disk fractional luminosity and best $\chi ^2 $ are listed in Table~\ref{tab : max_flux}. The maximum disk fractional luminosities are the highest allowed by the 3$\sigma$ lower limit on the visibilities. However, one has to keep in mind that this assumes fully resolved disks, which may not be the case, and there is a slim possibility 
that a very compact unresolved disk remains in the centre. 

For unresolved sources, the visibilities are compatible with 1.0 but can take slightly higher values because of the uncertainties. This is unphysical but numerically allowed by the calibration procedure. Because $f_{disk}$ is always positive, this can explain some higher $\chi^2$ values, such as for V709 CrA and TWA 3B.

These results are consistent with SED analysis of these objects. The SED of these stars shows an excess in the mid-infrared and sub-millimetre range, while the optical and near-infrared SED is compatible with a naked star without excess coming from either thermal emission or scattered light, suggesting that the inner parts of these disk have been cleared (at least down to undetectable levels). 

\begin{table}[t]
\caption{\label{tab : max_flux} Upper limits of disk luminosity in H-band for the unresolved stars.}
\centering
\begin{tabular}{l|cc}
\hline
\hline
Star&   Max F$_{disk} $/ F$_{tot} $ (\%) & $\chi ^2/(nb. vis.) $\\
\noalign{\smallskip}\hline\noalign{\smallskip}
V4046 Sgr& 3.16& 1.72 \\ 
TWA 07& 0.68& 0.68 \\ 
V709 CrA& 1.09& 3.32 \\ 
TWA 3B& 2.38& 1.91 \\ 
MY Lup& 6.31& 0.97 \\ 
FK Ser& 3.38& 0.57 \\ 
\hline
\end{tabular}
\end{table}

\section{Simple models to characterise the visibility profiles of resolved disks}
\label{sec:viscurv}

In addition to the unresolved sources (see \S \ref{sec : nondetec}) and to the resolved binary systems (see \S2.2  and \S7),
there are 13 (effectively) single targets in the sample that show a clear signature of the surrounding disk in their visibility profiles. Below we devise two simple models of disks in order to interpret and describe the visibility data of our sample. For this preliminary modelling, we considered pole-on models only because the closure phase signals are weak (usually compatible with zero as we can see from the closure phase profiles displayed Figure~\ref{app:data} and Table~\ref{tab : avgCP}), and we also neglected the eventual spread in the data caused by different projected baseline PA. The general properties are of interest here. In-depth model fitting that involves several other data sets will be presented elsewhere. 

\begin{table}[t]
\caption{\label{tab : avgCP} Mean closure phase, mean absolute value of the closure phase, and  mean error on the closure phase for each non-binary star with a resolved disk on our sample, in degrees. For all these object, the mean closure phase and the mean absolute closure phase are compatible with 0.}
\centering
\begin{tabular}{lccc}
\hline
\hline

 Star & $\overline{CP}$& $\overline{|CP|}$&$\overline{\sigma_{CP}}$\\
\noalign{\smallskip}\hline\noalign{\smallskip}

TW Hya     & -1.54&  1.87&  2.83\\
HT Lup     & -1.10&  2.13&  2.90\\
GQ Lup     & -1.70&  1.70&  2.10\\
RU Lup     & -2.09&  2.21&  2.11\\
RY Lup     & -2.58&  3.16&  3.58\\
V1149 Sco  & -0.01&  1.52&  3.08\\
AS 205A    & -0.98&  1.29&  1.59\\
V2129 Oph  & -0.67&  2.34&  3.25\\
V2508 Oph  & -4.05&  6.87&  6.86\\
V1121 Oph  & -0.27&  2.37&  5.10\\
S CrA      & -2.73&  2.73&  2.51\\
V380 Ori   & 1.17 &  4.20&  4.88\\
HN Lup     & 1.02 &  1.40&  4.94\\
\hline
\end{tabular}
\end{table}

\subsection{The ring model}
\label{subsec : ring model}
We first fit the visibility profiles with a ring model (hereafter the thermal model) discussed in \cite{Eisner03}. This model assumes that all the energy coming from the disk and the puffed-up inner rim is due to thermal emission. The model is made of a ring of constant brightness distribution and of width-to-radius ratio $w=0.18$. The squared visibility $V^2$ of the system (the star and the thermal ring) is
 \begin{equation}
V^2 =  \left ( \dfrac{1+f_{therm} V_{ring}}{1+f_{therm}}\right)^2 ,
\label{eq : Vring}
\end{equation}
where $V_{ring}$ is the visibility of the ring written as
\begin{equation}
V_{ring} = $ $\dfrac{1}{\pi  \theta_{in} q(2w+w^2) } \times  [ (1+w) J_1(2\pi (1+w)   \theta_{in}q) -J_1 (2\pi  \theta_{in}q) ].
\label{eq : Vring2}
\end{equation}
The visibility of the star is set to one and is unresolved. Here, $J_1$ is the Bessel function of the first kind, $ \theta_{in}$ the opening angle of the inner rim (radius), q=$\sqrt{u^2 + v^2}$ is the uv-distance, and $f_{therm}$ is the ratio of the thermal ring flux over the stellar flux: 
\begin{equation}
\label{eq : fexc}
f_{therm.} =  F_{disk}/F_{star} = F_{tot.}/F_{star} - 1
.\end{equation}

This model depends on two parameters, $\theta_{in}$ and $f_{therm}$. Here, $f_{therm}$ can be estimated if one knows the total and the stellar fluxes at H-band. 
\subsection{The composite model}
\label{subsec : scat model}

The second model is a refinement of the one presented above. It is motivated by the
PIONIER data obtained in compact configuration and the previous predictions made by \cite{pinte08}. Unfortunately, only three single targets could be observed at short baselines (HT Lup, RU Lup, and RY Lup), but they all show a rapid decrease in the visibility profiles at short baselines (<10M$\lambda$), indicating that an extended component is resolved, at least partially, and thus larger than R$\sim$3 au (at 150 pc).

 To identify the nature of this extended component we compare its size with the size of the emission zone produced by thermal emission only, at H-band. To do so we first calculated the size of the thermal emission zone of a flared disk \citep{Eisner04} without the contribution of a puffed-up inner rim. For a 10 L$_{\sun}$ star (brighter than all the stars in our survey except for V380 Ori), the emission region (where 99\% of the disk emission comes from)   is only $\sim$ 1.5au wide. Adding a puffed-up rim makes this size smaller because it concentrates an important part of the thermal emission closer to the centre, thus reducing the size of the emission region 
  This upper limit is half the size of the extended component. Lower luminosity values would lead to a correspondingly smaller zone. It seems reasonable to assume that the extended component is not due to thermal emission alone. Scattered light appears as a natural candidate. This assumption is also motivated by the high albedo of typical disk grains that can be up to 0.9 for small silicate particles.

To check the ability of scattered light to match the data, we build a composite model where thermal emission (see \S~\ref{subsec : ring model}) is combined with a scattered light component. For simplicity we
only consider isotropic scattering. To determine the radial dependence of the scattered flux, we determine the surface brightness of a power-law disk characterised by a flaring of its surface with exponent $\beta = 1.1$ (at $\tau = 1$), a scale height $H_o$, and an outer radius set equal to half of PIONIER's field of view\footnote{This flaring exponent is not the true flaring exponent of the disk, since the optical depth will decrease with the radius (owing to the radial decrease of the surface density), but calculations with different flaring values (from 1 to 1.25) lead to little differences in the results. In a similar way, $H_0$ is not the true scale height of the disk but the vertical distance up to $\tau$= 1. However, the value of $H_0$ has no influence on the calculations.}. 

The surface of a ring of width $dr$ at a radius $r$ of this disk can be written as  
\begin{equation}
dS (r) = 2\pi r \times\sqrt{ dr^2 + dH^2} = 2\pi r \times\sqrt{ 1 + (\alpha r^{\beta-1} ) ^2}dr
\label{eq : dS}
\end{equation}
where $H = H_0\left (r/r_ 0\right) ^\beta$ is the height of the disk at the radius $r$, $r_0$ the reference radius, and  $\alpha = H_0\beta/r_0^\beta$. 
This ring is illuminated by the star with an angle 
\begin{align}
 \rho &= \theta-\phi = tan^{-1} (dH/dr) - tan^{-1} (H/r) \\
   &= tan^{-1} ( \alpha r^{\beta-1}) -  tan^{-1} ( \alpha  r^{\beta-1}/\beta) 
\label{eq : rho}
\end{align}
where  $\theta$  is the slope of the ring, and  $\phi$  the angle between the ring, the star, and the midplane of the disk. Finally, the flux illuminating the ring is proportional to $1/(r^2+H^2),$ and the ring has an albedo $A$. Combining these terms with equations~\ref{eq : dS} and \ref{eq : rho}, the flux scattered by the disk at the radius r is 
\begin{equation}
dF (r) = 2\pi r\sqrt{ 1 + (\alpha r^{\beta-1} ) ^2}  \times \dfrac{A}{(r^2+H^2)}\times sin (\rho) dr.
\label{eq : dF}
\end{equation}
This ring flux is finally normalised to 1 and multiplied by the ratio $f_{scat} = F_{scat}$/$F_{\star}$ between the scattered light flux and the stellar flux,
\begin{equation}
df_{scat} (r) = \dfrac{dF (r) * f_{scat}}{\int dF (r)}.
\label{eq : dfscat}
\end{equation}

This normalisation has the advantage of being independent of the albedo and $H_0$ (as long as $\alpha<<1$ ). The visibility profile of this model is  
\begin{equation}
V^2 =  \left ( \dfrac{1+f_{therm} V_{ring}+ \int_{\theta_{in}}^{\theta_{out}}\left (df_{scat} V_{ring}    \right) }{1+f_{therm}+ f_{scat}}\right) ^2 .
\label{eq : V2scat}
\end{equation}
Each emission component shares the same inner radius, so this model has three free parameters ($\theta_{in}$, $f_{therm}$, and $f_{scat}$). 

Similar to the thermal model, the number of parameters can be reduced if one knows the disk-to-stellar flux ratio in H band $f_{exc.}$, which can derived using equation~\ref{eq : fexc} and replacing $f_{therm}$ by $f_{exc.}$.
%
In this case, and considering now that the excess flux is coming from both thermal emission and scattering, then $f_{therm}$ can be written $f_{exc.} - f_{scat}$ so the final free parameters of this model are $\theta_{in}$ and $f_{scat}$. In the section below we use these two models to fit the PIONIER data.

\section{Simple fits of the PIONIER data.}
\label{sec : mod data}
\subsection{Modelling the visibility profiles}
\begin{figure*}[ht]
\includegraphics[scale=0.80]{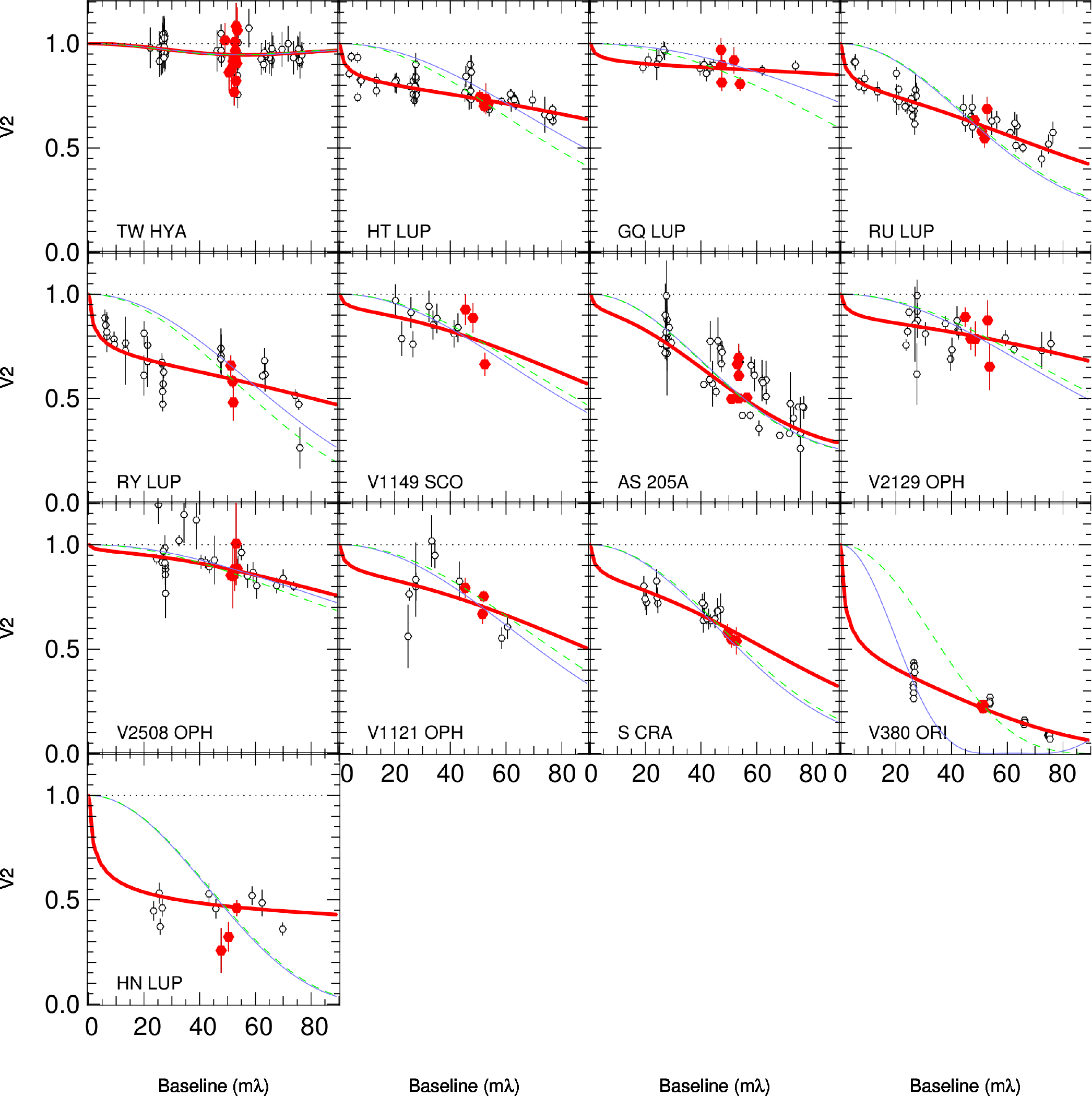}
\caption{\label{fig : sum_scat} Plots of the results form the visibility modelling with the thermal and composite models. 
For each star the black circles and vertical bars are the data points and error bars, and the dotted line represents V$^{2}=1 $. The red lines are the  fits with the composite model, and the blue lines are the fits with the thermal model. Fitting only the visibility point between $\sim$47 and $\sim$54 M$\lambda$ (roughly the baseline of the Keck Interferometer, see the red points and error bars) with the thermal model leads to similar results (green dashed line) for the majority of the stars as fitting the whole visibility profile.}
\end{figure*}

For each target, we list in Table \ref{tab : fit-res} the results from both models. The corresponding visibility profile plots are presented Figure  \ref{fig : sum_scat}.
Two stars have a published excess, $f_{exc.}$, at H-band: TW Hya \citep{Menu14} and S CrA \citep{Vural12}.  To estimate this excess for the rest of the targets, we performed a spectral decomposition by fitting the visible part of the SED with a Kurucz model, with the effective temperature and luminosity fixed to the values found in the literature (see Table~\ref{tab : dist_star}).  Then $f_{exc.}$ is derived using eq.~\ref{eq : fexc}.
The resulting excesses are presented Table~\ref{tab : exces}.

\begin{table}[t]
\caption{\label{tab : exces}  Derived values of $f_{exc}$ for the 13 resolved disks of our survey. $f_{exc}$ is defined as $f_{disk}/f_{star}$ and has been either taken from the literature (TW Hya and S CrA) or estimated by spectral deconvolution.}
\centering
\begin{tabular}{lc}
\hline\hline\noalign{\smallskip}
Star&$f_{exc}$\\
\noalign{\smallskip}\hline\noalign{\smallskip}
TW Hya          & 1.03    $\pm0.01$     \\
HT Lup          & 1.61    $\pm0.07$     \\
HN Lup          & 2.53    $\pm0.14$\\
GQ Lup          & 2.18    $\pm0.07$     \\
RU Lup          & 1.66    $\pm0.09$      \\
RY Lup          & 1.98    $\pm0.28$      \\
V1149 Sco       & 1.21    $\pm0.11$     \\
AS205 A         & 1.58    $\pm0.05$      \\
S CrA           & 2.50    $\pm0.15$      \\
V2129 Oph       & 1.10    $\pm0.20$     \\
V2508 Oph       & 1.24    $\pm0.18$      \\
V1121 Oph       & 1.49    $\pm0.16$      \\
\noalign{\smallskip}\hline\noalign{\smallskip}
\end{tabular}
\end{table}

The range of validity for  $R_{in}$ and $f_{therm}$ was derived by computing the $\chi^2$ map of the model results, then deriving the associated marginalised probabilities 
  \begin{equation}
  \label{eq:pft}
  p_{composite}(f_{therm}) \propto \sum_{R_{in} = 0}^{\infty} e^{(-\chi^2{(R_{in}, f_{therm})}/2)}
  \end{equation}
  and
  \begin{equation}
  \label{eq:pri}
  p_{composite}(R_{in}) \propto \sum_{f_{therm} = 0}^{f_{tot}} e^{-(\chi^2{(R_{in}, f_{therm})}/2)}
  \end{equation}
  and defining a 68\% confidence interval around the best model along each axis (i.e, for $R_{in}$and $f_{therm}$). Equations~\ref{eq:pft} and \ref{eq:pri} are valid because $f_{therm}$ and R$_{in}$ are sampled uniformly in the models.  The validity range of the thermal model was derived similarly, computing the $\chi^2$ of each model as a function of $ R_{in}$, then the associated probability 
   \begin{equation}
  p_{therm}(R_{in}) \propto  e^{-\chi^2(R_{in})/2},
  \end{equation}
   and finally defining a 68\% confidence interval around the best model. 
   
We caution that the thermal model provides poorer fits to several of the data sets (see Fig. \ref{fig : sum_scat}).
The average value for the thermal models' reduced $\chi^2$ is 9.2 and is 3.1 for the composite model. 
The median value of the thermal model's  $\chi^2_{red}$  is 4.9 (with values up to 25 as presented  Table~\ref{tab : fit-res}, here we neglect V380 Ori and HN Lup that may be associated with envelopes, see below). In this case, meaning with poor models, and although error bars and the validity range can be formally calculated, the exercice leads to validity ranges for $R_{in}$ that are not representative. We list them for completeness in Table~\ref{tab : fit-res} but caution that these errors are not reliable for the thermal model. 


\begin{table*}[t]
\caption{\label{tab : fit-res} Inner rim size, thermal excess, scattered excess, and  $\chi^2_{red}$ for the composite model (left part), and the thermal model (right part). f$_{scat}$ is the difference between the total excess flux (estimated from SED analysis) and f$_{therm.}$. 
}
\centering
\begin{tabular}{lrrrr|rrr}
\hline\hline\noalign{\smallskip}
Star&\multicolumn{4}{c}{Composite model}&\multicolumn{3}{|c}{Thermal model}\\
\noalign{\smallskip}\hline\noalign{\smallskip}
 & $R_{in}$[au]& f$_{therm}$& f$_{scat}$& $\chi^2_{red}$& $R_{in}$[au]& f$_{therm}$&  $\chi^2_{red}$\\
\noalign{\smallskip}\hline\noalign{\smallskip}
 TW Hya         & 0.111$^{+0.000}_{-0.028}$     & 0.02$^{+0.00}_{-0.01}$         & 0.01  & 0.89          & 0.120$^{+0.019}_{-0.018}$     & 0.03   & 0.88 \\ 
 HT Lup         & 0.055$^{+0.004}_{-0.004}$     & 0.42$^{+0.01}_{-0.01}$         & 0.22  & 2.21          & 0.100$^{+0.000}_{-0.000}$     & 0.64   & 16.09 \\ 
 GQ Lup         & 0.041$^{+0.013}_{-0.034}$     & 0.12$^{+0.01}_{-0.02}$         & 0.08  & 1.36          & 0.107$^{+0.002}_{-0.002}$     & 0.20   & 3.38 \\ 
 RU Lup         & 0.102$^{+0.003}_{-0.004}$     & 0.41$^{+0.01}_{-0.01}$         & 0.25  & 2.20          & 0.149$^{+0.001}_{-0.001}$     & 0.66   & 22.79 \\ 
 RY Lup         & 0.065$^{+0.006}_{-0.006}$     & 0.52$^{+0.02}_{-0.02}$         & 0.41  & 3.83          & 0.119$^{+0.001}_{-0.001}$     & 0.93   & 25.28 \\ 
 V1149 Sco      & 0.053$^{+0.013}_{-0.021}$     & 0.51$^{+0.03}_{-0.07}$         & 0.09  & 1.72          & 0.070$^{+0.003}_{-0.003}$     & 0.59   & 1.82 \\ 
 AS 205A        & 0.176$^{+0.002}_{-0.001}$     & 0.49$^{+0.01}_{-0.01}$         & 0.09  & 13.79         & 0.182$^{+0.000}_{-0.000}$     & 0.58   & 16.19 \\ 
 V2129 Oph      & 0.063$^{+0.009}_{-0.010}$     & 0.23$^{+0.01}_{-0.02}$         & 0.11  & 2.55          & 0.104$^{+0.001}_{-0.001}$     & 0.34   & 5.77 \\ 
 V2508 Oph      & 0.112$^{+0.008}_{-0.008}$     & 0.09$^{+0.01}_{-0.01}$         & 0.02  & 1.52          & 0.123$^{+0.003}_{-0.003}$     & 0.11   & 1.62 \\ 
 V1121 Oph      & 0.092$^{+0.012}_{-0.015}$     & 0.29$^{+0.03}_{-0.04}$         & 0.13  & 3.13          & 0.123$^{+0.002}_{-0.002}$     & 0.42   & 4.46 \\ 
 S CrA          & 0.078$^{+0.007}_{-0.008}$     & 1.20$^{+0.04}_{-0.06}$         & 0.30  & 0.69          & 0.109$^{+0.001}_{-0.001}$     & 1.50   & 3.04 \\ 
 V380 Ori       & 0.327$^{+0.003}_{-0.003}$     & 1.23$^{+0.02}_{-0.02}$         & 1.47  & 9.62          & 0.832$^{+0.001}_{-0.001}$     & 2.70   & 156.97 \\ 
 HN Lup         & 0.005$^{+0.024}_{-0.005}$     & 0.34$^{+0.36}_{--0.12}$         & 1.07  & 3.03          & 0.151$^{+0.001}_{-0.001}$     & 1.53   & 36.06 \\ 
\end{tabular}
\end{table*}
  
For 5 of the 13 resolved stars, both models lead to similar inner radii estimations. Several explanations are possible:
\begin{itemize}
\item  These disks may be flat or made of grains with low albedo, resulting in less scattering. V2508 Oph and AS 205A might fall into this 
category, because the thermal model can fit the visibility data of their well-resolved, luminous disks.
\item The uv coverage may be too poor (as for V1149 Sco) or may show a large dispersion in the visibility measurements (as for V1121 Oph),
resulting in poorly constrained models, 
\item The disk surface brightness may be low (as for TW Hya). As a consequence, the visibility drop-off is small enough that, even with short baselines measurements, both models lead to similar results.
\end{itemize}

For the remaining eight stars -- V380 Ori, HT Lup, HN Lup, GQ Lup, RU Lup, RY Lup, V2129 Oph, and  S CrA -- applying the thermal model results in poorer fits of the visibility profiles, while the composite model provides a better match (see Figure~\ref{fig : sum_scat}). Interestingly, we note that the composite model (with scattering) does much better for RU Lup, HT Lup, and RY Lup, the only three disk targets for which short baseline measurements are available. Because scattered light is expected to produce clear signatures at short baselines, it will be important to verify the solidity of this trend with more data obtained in compact VLTI configurations. 

We also note that the values of f$_{scat}$ listed in Table \ref{tab : fit-res} vary from 0.0 (dominated by pure thermal emission) to above {f$_{therm}$} (scattered light is significant). V380 Ori and HN Lup have the highest values of f$_{scat}$, but envelopes may contaminate the results. Excluding them for safety, the values range from 0 to $\sim 40$\% of the total disk flux. This is a wide range that indicates that disks are likely to be different from one another, either in shape or content. The current data does not allow detailed image reconstruction. However, more interferometric data and detailed modelling of individual sources, to be performed elsewhere and adding information from several other data sets, will be useful for exploring the shape and content of the inner disks around T Tauri stars further.   

A few disk targets are worth a special note, in particular because an extended envelope may affect the interferometric measurements, as mentioned above. V380 Ori is associated with a massive (7.6$_{-2.8}^{+4.4}M_{\sun}$) envelope \citep{Liu11}. The envelope is resolved with the shortest baselines in our observations. It is responsible for the large visibility decrease at short baselines. In this case, both disk models are inappropriate, since incomplete because of the envelope. The data for HN Lup are scarcer, but SED-fitting permits estimating the excess flux in the H band to roughly about 1.5 times the photospheric flux. This is a huge excess flux that hints at the presence of a possible envelope around HN Lup as well. There is a hint, from extended CO line emission, that an envelope may also be associated with V1149 Sco \citep{Dent05}. 

A comparison of our values of $R_{in}$ with previous estimates from the literature is also interesting. We find that the inferred inner radius for S CrA with the thermal model is compatible with the previous estimates from \cite{Vural12}. The radii of AS 205 A are also consistent with values found by \cite{Eisner05} with a puffed-up rim, flared-disk model. However, our inferred radii of TW Hya are slightly smaller than the one found in \cite{Menu14} (0.11 au versus 0.3 au). All others are new estimations.

We note that our thermal model leads to similar inner rim radii regardless of whether we fit the whole visibility profile or only the data between 80 and 90 meters (i.e., the  Keck Interferometer and the baseline range used in previous studies by \cite{Akeson05b} and \cite{Eisner07}).  This is serendipitous, because having $\sim 80$ m baselines is close to the average baseline range in our coverage. Nevertheless, a comparison remains useful with previous studies made with thermal models and with data for this specific restricted range of baselines (see \cite{pinte08} and references therein).

\subsection{Comparison with the dust sublimation radius}
\begin{figure}[h]
\includegraphics[width=9cm]{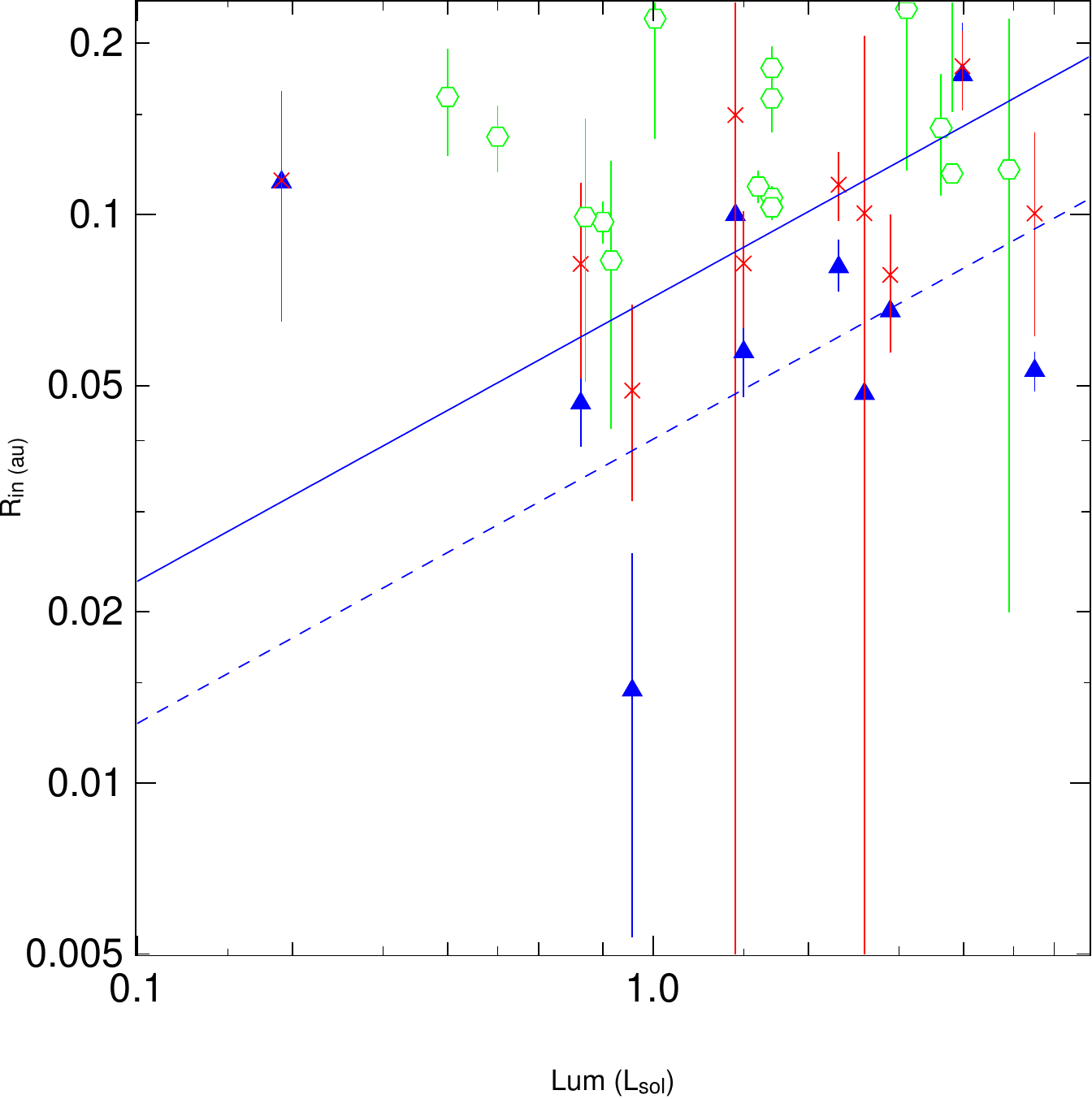}
\caption{\label{fig :models RinLum} Calculated inner rim for all the single, resolved stars of our survey (except V380 Ori and HN Lup) with the composite model (blue triangles and error bars) and the thermal model (red crosses and error bars). Radii calculated with thermal models from the literature (Pinte et al. 2008) are also presented by green circles.  The blue lines represent the sublimation radius as a function of the stellar  luminosity for grains sublimating at 1500 K (straight line) and 2000 K (dashed line).  All estimations assume a pole-on configuration.}
\end{figure}

We present in Fig.~\ref{fig :models RinLum} a comparison of the inner radii derived from the two models with the sublimation radius estimated for an optically thin disk. See the straight and dashed lines in the figure~\ref{fig :models RinLum}, which are calculated with the following prescription: 
\begin{equation}
R_{sublim}= \sqrt{(1+H_{in}/R_{in})\dfrac{L_{star}+L_{acc}}{4\pi\sigma T_{sub}^4}},
\label{eq : Rsub}
\end{equation}
where we adopted  H$_{in}$/R$_{in}$ (0.1) and $T_{sub}$ (1500 and 2000 K) for direct comparison with \cite{Eisner07}.
In addition, we plot the radii calculated by \cite{pinte08} using only a thernal ring and archival data from Keck and PTI. 

The radii from the thermal model are compatible with the dust sublimation radius, except for TW Hya.  Not surprisingly perhaps (see \cite{pinte08}), the radii calculated with the composite model are smaller than their thermal model equivalent. They are also consistent with an inner rim at the sublimation radius, assuming a 1500-2000K dust sublimation temperature. 

Two stars have an inferred radius that is small and below the 2000K dust sublimation curve: RY Lup and HT Lup. Their small inner rims can be accounted for by large projection effects  of high inclinations. These effects have to be taken into account properly to extract better parameters.  


As presented in this section, geometrical models are sufficient to highlight the general properties of T Tauri disks and, in particular, here the presence (or absence) of a significant contribution from (extended) scattered light. These simple models also highlight the possibility that the inner regions of each disk is different from target to target and that envelopes may also contaminate the interferometric signal if not taken into account properly.  
Full radiative transfer modelling is required to derive more precise parameters about the structure and composition of the inner disks. Such modelling is currently underway for all the stars with disks discussed above. These models will be presented elsewhere.

\section{Binaries detected in the survey}
\label{sec : bin}

The companions to four stars were detected during this survey, three of which were already known from spectroscopy. This is the first time the binary is resolved spatially for all four objects. Table~\ref{tab : binairesnew} contains results for WW Cha, V1000 Sco, and TWA 3A: the date of observation, the separation and position angle of each companion, and the flux ratio between companion and primary.  AK Sco is discussed in detail below, and its orbital parameters are presented in Table~\ref{tab:aksco_bestfit}.

\begin{table*}[t]
\caption{\label{tab : binairesnew} Date of observation, separation, position angle, astrometric errors, and flux ratio of the observed binaries. The astrometric errors are the semi major axis \textbf{a}, the semi minor axis \textbf{b,} and the orientation of the error ellipse, $\theta$, around the best fit position of the companion.  The flux ratio of TWA 3A is poorly constrained by the observations.
}
\centering
\begin{tabular}{cccccccc}

\hline
Star &  Date of observation& Sep.(mas)& P.A. ($\degr$)&a(mas)&b(mas)&$\theta$($\degr$)&  $F _{ (comp) }/F _{ \star }   $ \\
\hline
\hline
&&&&\\
WW Cha          &2012-07-02&4.73&-140&0.57&0.35&22      &0.62 \\
                        &2012-03-06&6.18&-48.8&0.98&0.40&117&0.62 \\
                        &2011-02-10&6.44&173.4&0.50&0.18&150&0.62        
                        \smallskip\\ 
V1000 Sco       &2012-07-17&4.33&105.7&0.40&0.19&164&0.34        \\
                        &2012-08-19&\multicolumn{6}{c}{- - - - - - - - - - multiple local minima - - - - - - - - - -}     
                        \smallskip\\
TWA 3A          &2011-02-09&3.51&108.1&0.57&0.43&1      &0.76 \\
\hline

&&\\
\end{tabular}

\end{table*}

\subsection{AK Sco}
\label{sec:AKSco}
AK Sco is a double-line spectroscopic binary. We observed it during seven nights between August 2011 and July 2013 (Table~\ref{app:data}) with the VLTI configured with long (6 nights) and short (one night in July 2013) baselines. Data were spectrally dispersed over one or three spectral channels across the H band. The dataset reveals the presence of the central binary plus an extended surrounding environment. 

\begin{figure}
  \centering
  \includegraphics[width=0.45\textwidth]{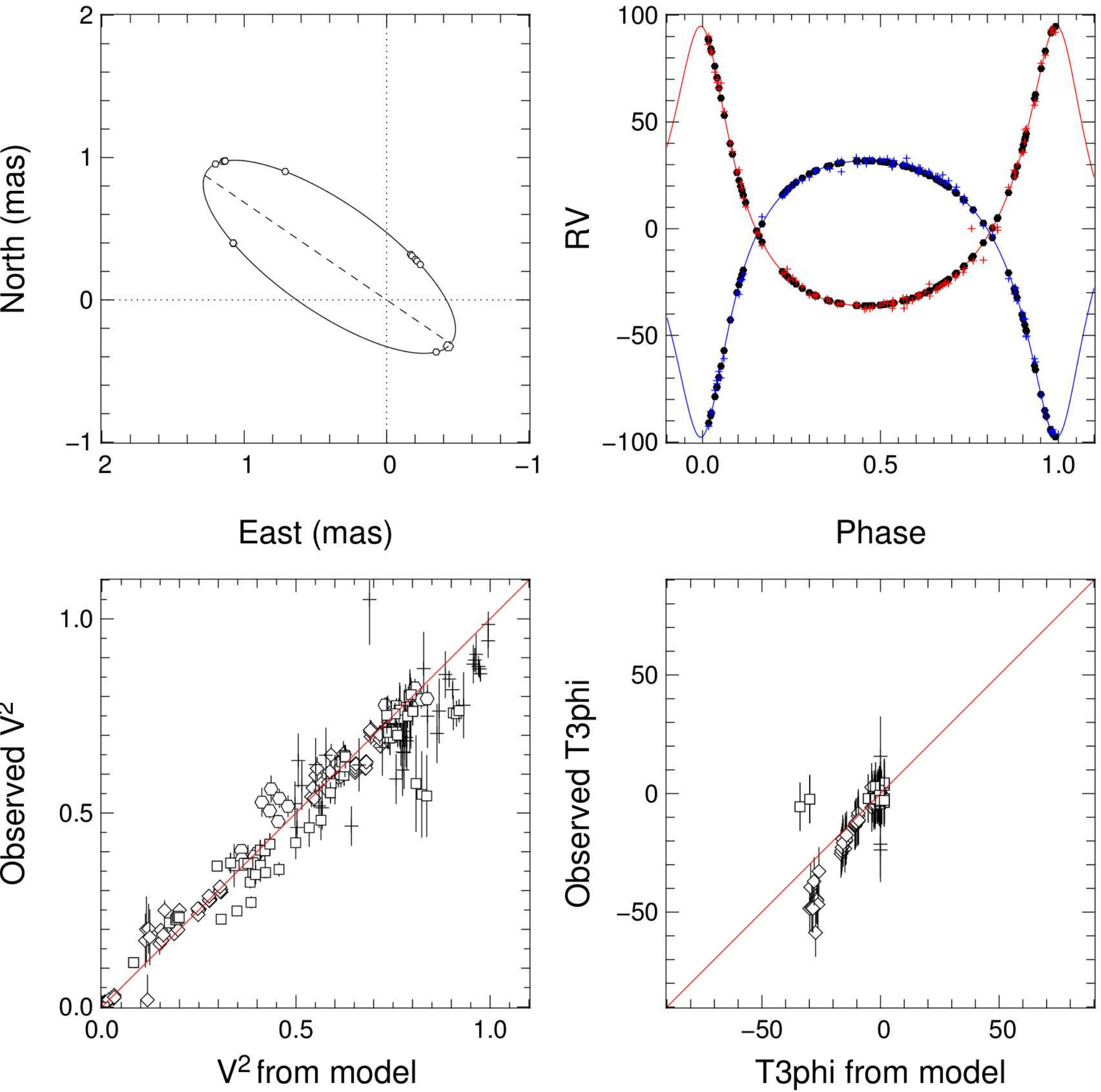}\vspace{1cm}
  \includegraphics[width=0.45\textwidth]{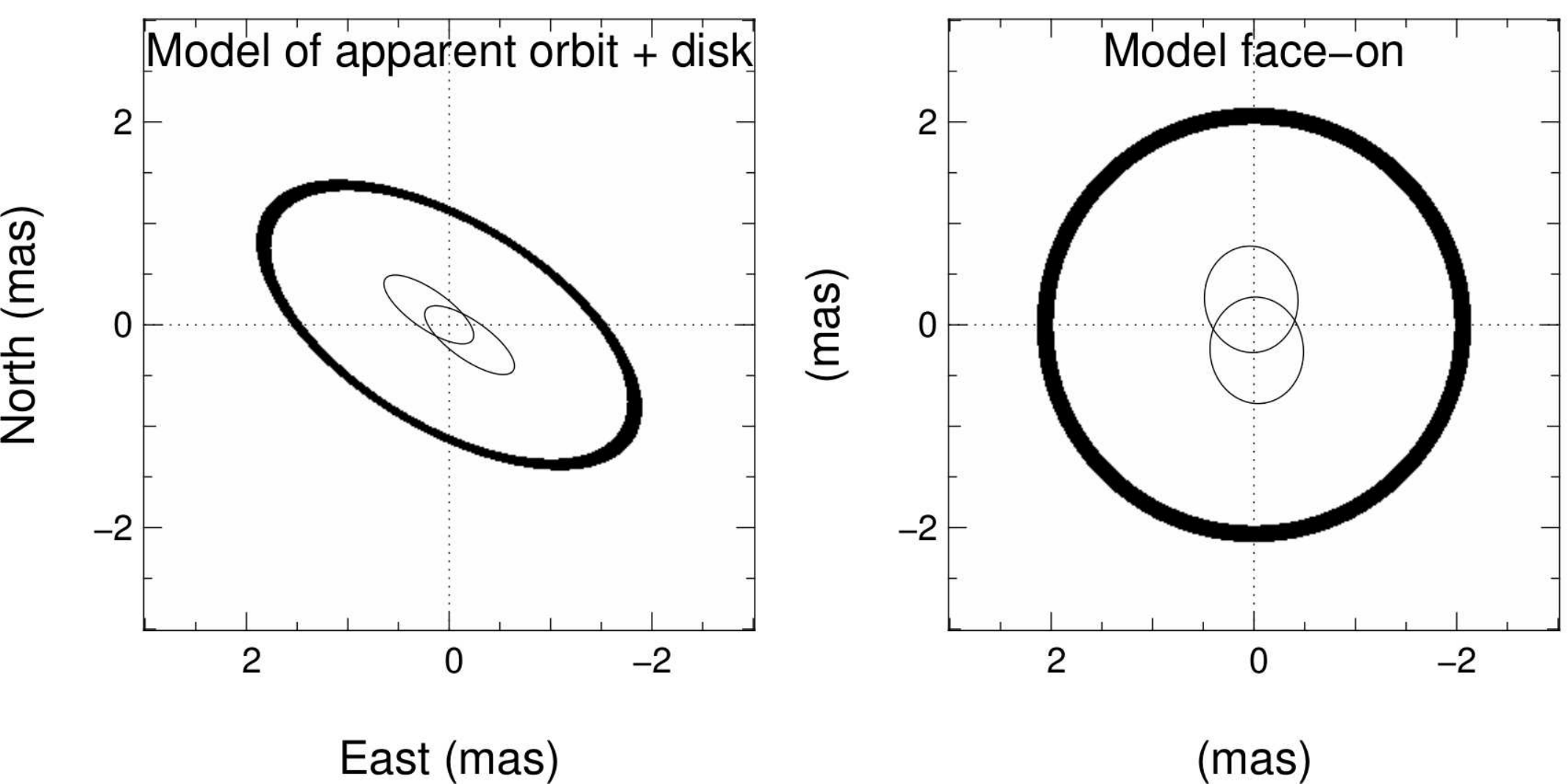}
  \caption{\label{fig:fit_aksco}Best fit of AK Sco's radial velocities and interferometric data with the binary+ring model. From top left to bottom right:\newline \textit{upper left}: Apparent trace of the best-fit orbit of the binary (primary at 0, 0) with the time of observation; \textit{upper right}: radial velocity from Alencar with the best-fit orbit;\newline \textit{middle left}: comparison between model and PIONIER  observation for V2; \textit{middle right}: comparison between model and PIONIER observations for closure phases; \newline\textit{lower left}: full model of disk + binary as projected on the sky; \textit{lower right}: full model of disk + binary as seen pole-on. }
  
\end{figure}

A first fit of interferometric data and radial velocities (from \cite{alencar03}) with a model consisting of a binary without circumstellar material leads to poor results, with the reduced $\chi^2 $ of the fit being $\chi^2_r=18 $ and the residuals showing the signature of an extended environment. A second model consisting of a binary surrounded by a narrow ring (to mimic the inner edge of the disk) is able to fit radial velocities, visibilities, and closure phase more successfully ( $\chi^2_r\approx2 $). The visual agreement between the interferometric measurements and the second model is also more convincing, without systematic residuals at low or high spatial frequencies. The best-fit parameters are summarised in Table~\ref{tab:aksco_bestfit} and plotted in Figure~\ref{fig:fit_aksco}.  

The radial velocities constrain the orbital elements $P $, $T $, $e $, $\omega $, $K_a $, $K_b $, and $\gamma $ very well.  The addition of the interferometric observations constrained the position angle of the ascending node and inclination, $\Omega$ and $i$, respectively, as well as the size of the apparent orbit.

The inclination of the binary, hence the stellar masses, are compatible with those found in  \cite{alencar03}. The diameter of both stars is less than 0.15 mas and cannot be resolved with the longest baseline of the VLTI, so  we consider the star to be unresolved in the fitting process.  The distance is constrained with the period, masses, and apparent size of the orbit of the binary. It is consistent at 1.5$\sigma$ level with the value from \cite{VanLeeuwen07} (d $=102_{-17}^{+26}$pc), while being more precise.
 
Interferometric observations also constrain the parameters of the circumbinary disk (major axis $w_{disk} $, inclination $i_{disk} $, position angle $\Omega_{disk} $, and the fractional flux of the disk $f_{disk} $). Interestingly, at one observation epoch, the binary was at closest apparent separation and was nearly unresolved. At that time, the interferometric signal was coming mostly from the disk.  

The number of interferometric data points and the spatial resolution are not enough to disentangle the exact nature of the environment.  A model of a binary plus an extended environment is able to reproduce the signature of the environment. This environment can be either an inclined ring, a uniform disk, or a Gaussian disk. Depending on the model considered, its diameter varies between $3.5\mas $ to $6\mas$, the values listed in Table~\ref{tab:aksco_bestfit} are for the ring model. The uncertainties reported in Table~\ref{tab:aksco_bestfit} are based on the statistical errors of the observations alone and not on the modelling uncertainty. However, the flux ratio between the environment (circumbinary disk or ring) and the central binary is robust, and the flux contribution from the disk ( f$_{disk}=0.2 $) is compatible with the excess flux in H band.

Our data set cannot distinguish between $i_{disk}=60\deg $ and $i_{disk}=180-60\deg $, no information on the disk rotation sense is available. However, the position angle of the disk is compatible with the binary one, and one of the two values of  $i_{disk}$ matches the inclination of the orbital plane. Based on that, it seems reasonable to conclude that the two are coplanar, although we acknowledge that only a spectro-astrometric measurements of a sense of rotation of the disk would confirm this definitively. 

%

%

\begin{table}[t]
\caption{Best fit orbital elements and related physical parameters of AK Sco}
\centering
\begin{tabular*}{0.95\columnwidth}{llll}
\hline\hline\noalign{\smallskip}
Parameter & Value  & Uncertainty & Unit/definition\\
\noalign{\smallskip}\hline\noalign{\smallskip}
$\gamma$ & $-1.97$  & $0.5$ & km\,s$^{-1}$\\
\Ka &  $64.7$ & $0.9$ & $\mbox{km\,s}^{-1}$ \\
\Kb &  $65.5$ & $0.9$ & $\mbox{km\,s}^{-1}$ \\
$t_\mathrm{p}$ & $46654.410$ & $0.02$ & {\tiny JD-2400000}\\
$P$ &  $13.609$ & $0.001$ & days \\
$e$ &  $0.47$ & $0.01$ & \\
$i$ &  $115$ & $3$ & $\deg$ \\
$\Omega$ &  $48$ & $3$ & $\deg$\\
$\omega$ &  $186$& $2$ &$\deg$\\
\Aap &  $1.11$ & $0.04$ & mas \\
$f$ & $0.81$ & $0.06$ & H-band flux ratio \\
\noalign{\smallskip}\hline\noalign{\smallskip}
f$_{disk}$ & $0.18$  & $0.03$ & flux.env \\
$i_{disk}$ & $121$ & $8$ & $\deg$ \\
$w_{disk}$ & $4.1$ & $0.3$ & $\mas$\\
$\Omega_{disk}$ & $47$ & $10 $ & $\deg$ \\
\noalign{\smallskip}\hline\noalign{\smallskip}
$d$ &  $141$ & $7$ &pc \\
\Ma &  $1.41$& $0.08$ &$\Msun$ \\
\Mb &  $1.39$& $0.08$ & $\Msun$ \\
\noalign{\smallskip}\hline
\end{tabular*}
\label{tab:aksco_bestfit}
\end{table}

\subsection{V1000 Sco}
V1000 Sco is a spectroscopic binary, some of its parameters ( $P $,$e$,  $\omega $,  $K_a $, and $\gamma $) have been previously measured by \cite{Mathieu89}. V1000 Sco has been observed twice during our campaigns. The first observation set is of good quality, and the object was observed twice during that night. The companion position and flux can be modelled without ambiguity. The resulting separation and flux ratio are listed Table \ref{tab : binaires}. This binary was observed again 33 days later. The data is of lower quality, and the modelling of the companion parameters led to numerous local minima. Further observations are being collected to fully constrain the orbital parameters.

Because of the possible presence of excess, uncertain system age, and systematic uncertainties with evolutionary models, it is not possible to reliably convert the H-band flux of the companion into a mass. More multi-wavelength and/or spectroscopic information on the companion is needed to assess the binary mass ratio. This also applies to TWA 3A and WW Cha below.

\subsection{TWA 3A}
 TWA 3A was observed once on February 2011, and the interferometric data are in good agreement with a binary model, with a binary separation of 3.51 mas and flux ratio of 0.76. As for V1000 Sco and WW Cha, further observations will be needed to constrain this system. 

\subsection{WW Cha} 
\begin{figure}[t]
\includegraphics[width=9cm]{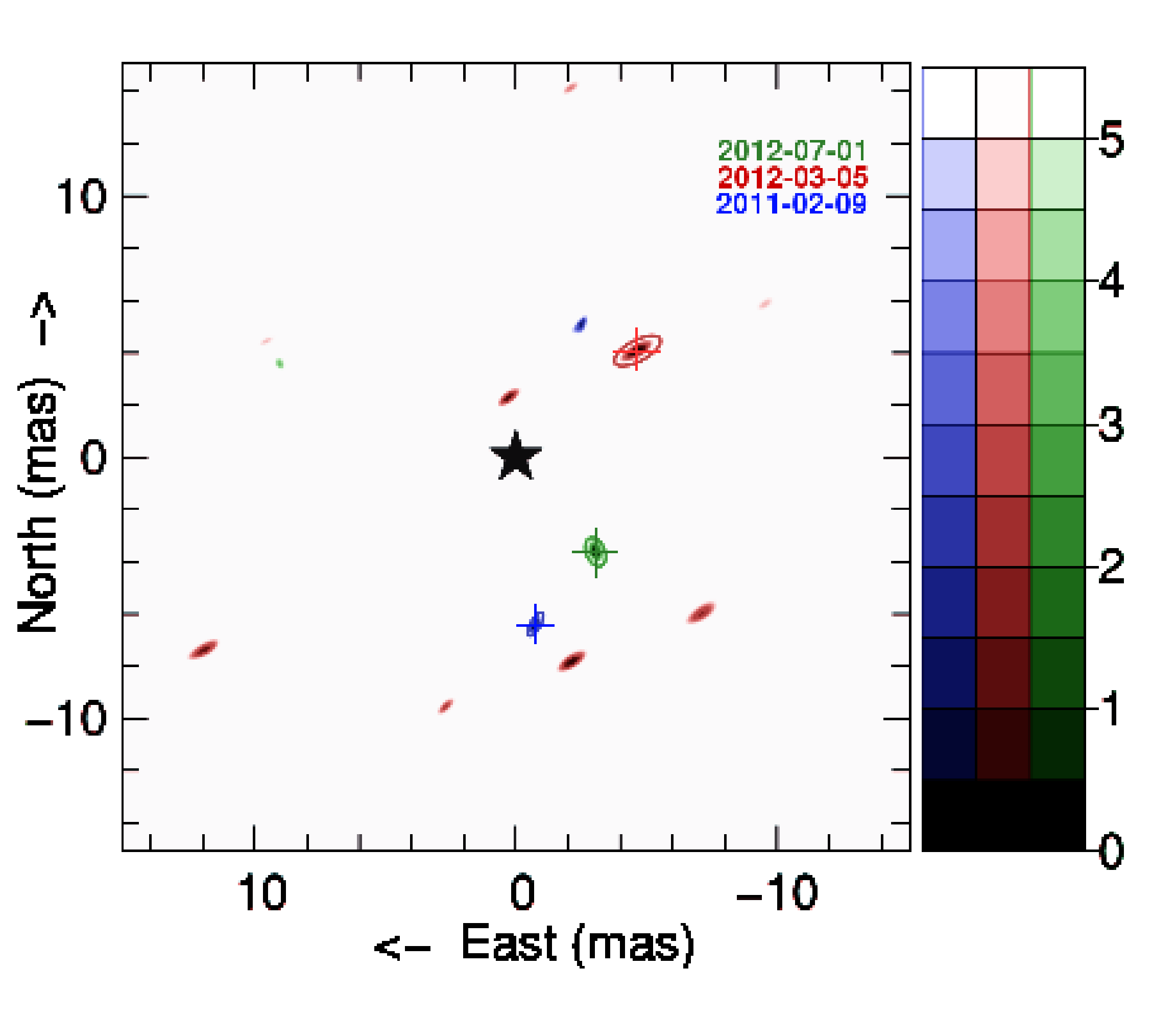}
\caption{\label{fit_wwcha} $\chi^2$ map for the binary model fitting of WW Cha for the observations of February 2011 (blue), March 2012 (red), and July 2012 (green) (for each colour, darker tones mean better $\chi^2 $). The primary is represented by the central black star, while the best fit position of WW Cha companion (i.e., the minimum  $\chi^2$) is denoted by a cross. Observations of March 2012 produced data of lower quality, leading to numerous local minima in this map.}
\end{figure}

WW Cha has been observed three times between February 2011 and July 2012. Each observation shows evidence for binarity (high visibility and closure phases variations, see Appendix \ref{app:images}). We performed interferometric fitting on these data with a model consisting of a binary and a fully resolved component (a disk or envelope). The amount of data is not sufficient to constrain the orbital parameters.  While the data of March 2012 is of lower quality than the two others (giving multiple local minima for the position of the companion), the fitting of all three data sets gives similar results for the companion's flux (0.6 time the flux of the primary) and the extended component (accounting for 12\% of the total emission). Figure~\ref{fit_wwcha} shows the apparent movement of the companion. This motion is not compatible with that of a background star, and it suggests that the two objects are bound.

\section{Summary and conclusion}
\label{sec : fin}

We observed 23 young stars with PIONIER at the VLTI, 21 T Tauri stars, and 2 Herbig Ae stars.
Thirteen of these stars have a visibility profile showing the clear signature of a circumstellar disk.  We fitted the data with two simple models: a thermal ring model and a more complex composite model consisting of the  thermal ring model and an additional large scale component. The composite model
significantly improves the quality of the fits in eight cases, especially the shape of the visibility curves at short baselines. These results strongly support the previous suggestion by \cite{pinte08} that the extended component is produced by scattered light.  For the five other cases, the fits from the two models are of similar quality . 

All in all, the model results indicate that the amount of scattered flux differs from disk to disk, as does the amount of thermal emission from the inner rim. This indicates that each disk is probably different from all others in its fine details. Clearly more advanced modelling is needed, including more constraints from other data sets, to derive the exact shape and content of each inner disks. This will be presented elsewhere.    

The size of the inner disks (or the inner rims) were also estimated for disk sources.  All the inner disk radii derived from the composite model are smaller than their counterparts from the thermal model. These radii are consistent with the radii expected for a dust sublimation temperature of $\sim$1500K, even for the faintest stars of our survey, extending the previous results of \cite{pinte08}.  Three objects (HN Lup, V1149 Sco, and V380 Ori) appear to have a remnant envelope and cannot be correctly represented with either models.

Improving our knowledge of the geometry and content of the inner regions of protoplanetary disks is necessary to understand how planets can form and how their migration can be halted. It is also necessary to understand the topology of the accretion process. T Tauri stars are less luminous and cooler (redder) than Herbig stars. Our results highlight the need to include a complete treatment of the radiative transfer (including scattered light) to interpret their NIR interferometric signatures.  They also highlight that with partial uv coverage, a very detailed estimation of the shape of the inner disk remains difficult. We stress that, ideally, future observations should include better uv coverage and simultaneous photometry. 


Another result of this survey is the discovery of a new binary: WW Cha. Its interferometric data can be modelled with a binary system and resolved  emission. The companion is seen at a different position during each observation, but further observations will be needed to fully constrain the orbital parameters of this milli-arcsecond binary. We also re-observed and spatially resolved the known binaries V1000 Sco, TWA 3A, and AK Sco. For V1000 Sco and TWA 3A, the separations, position angles, and flux ratios of the systems were derived. For AK Sco, both the binary system and the inner edge of the circumbinary disk are resolved. The orbital parameters and the distance to the system were derived. This model reproduces both the radial velocity measurements and the interferometric data. However, the exact shape of the circumbinary environment cannot be constrained yet from our observations. More data points and longer baselines are needed.

\begin{acknowledgements}
PIONIER is funded by the Universit\'e Joseph Fourier (UJF, Grenoble) through its Poles TUNES and SMING, the Institut de Plan\'etologie et d'Astrophysique de Grenoble, the ``Agence Nationale pour la Recherche'' with the programme ANR EXOZODI, and the Institut National des Science de l'Univers (INSU) via the ``Programme National de Physique Stellaire'' and ``Programme National de Plan\'etologie''. The authors want to warmly thank all the people involved in the VLTI project. This work is based on observations made with the ESO telescopes. It made use of the Smithsonian/NASA Astrophysics Data System (ADS) and of the Centre de Donn\'ees astronomiques de Strasbourg (CDS). All calculations and graphics were performed with the open source software \texttt{Yorick}. 
We acknowledge funding from the European Commission's 7$^{th}$ framework programme (EC FP7) under grant agreement No. 284405 (DIANA) and contract PERG06-GA-2009-256513 and also from Agence Nationale pour la Recherche (ANR) of France under contract
ANR-2010-JCJC-0504-01. 
FM,  SC and MS acknowledge support from Millennium Science Initiative, Chilean Ministry of Economy: Nucleus P10-022-F.
This research has made use of the Simbad database operated at the CDS, Strasbourg, France, and the Jean-Marie Mariotti Center \texttt{ASPRO} and \texttt{LITpro} services co-developed by CRAL, IPAG, and FIZEAU. We  thank the anonymous referee for her/his suggestions for improvement.\\
 \end{acknowledgements}

\bibliographystyle{aa} 
\bibliography{biblio2} 

\begin{thebibliography}{59}
\expandafter\ifx\csname natexlab\endcsname\relax\def\natexlab#1{#1}\fi

\bibitem[{{Akeson} {et~al.}(2005){Akeson}, {Boden}, {Monnier}, {Millan-Gabet},
  {Beichman}, {Beletic}, {Calvet}, {Hartmann}, {Hillenbrand}, {Koresko},
  {Sargent}, \& {Tannirkulam}}]{Akeson05b}
{Akeson}, R.~L., {Boden}, A.~F., {Monnier}, J.~D., {et~al.} 2005, \apj, 635,
  1173

\bibitem[{{Alecian} {et~al.}(2009){Alecian}, {Wade}, {Catala}, {Bagnulo},
  {B{\"o}hm}, {Bouret}, {Donati}, {Folsom}, {Grunhut}, \&
  {Landstreet}}]{Alecian09}
{Alecian}, E., {Wade}, G.~A., {Catala}, C., {et~al.} 2009, \mnras, 400, 354

\bibitem[{{Alecian} {et~al.}(2013){Alecian}, {Wade}, {Catala}, {Grunhut},
  {Landstreet}, {Bagnulo}, {B{\"o}hm}, {Folsom}, {Marsden}, \&
  {Waite}}]{Alecian13}
{Alecian}, E., {Wade}, G.~A., {Catala}, C., {et~al.} 2013, \mnras, 429, 1001

\bibitem[{{Alencar} {et~al.}(2003){Alencar}, {Melo}, {Dullemond}, {Andersen},
  {Batalha}, {Vaz}, \& {Mathieu}}]{alencar03}
{Alencar}, S.~H.~P., {Melo}, C.~H.~F., {Dullemond}, C.~P., {et~al.} 2003, \aap,
  409, 1037

\bibitem[{{Andersen} {et~al.}(1989){Andersen}, {Lindgren}, {Hazen}, \&
  {Mayor}}]{andersen89}
{Andersen}, J., {Lindgren}, H., {Hazen}, M.~L., \& {Mayor}, M. 1989, The
  Messenger, 55, 45

\bibitem[{{Andrews} {et~al.}(2010){Andrews}, {Wilner}, {Hughes}, {Qi}, \&
  {Dullemond}}]{Andrews10}
{Andrews}, S.~M., {Wilner}, D.~J., {Hughes}, A.~M., {Qi}, C., \& {Dullemond},
  C.~P. 2010, \apj, 723, 1241

\bibitem[{{Artemenko} {et~al.}(2012){Artemenko}, {Grankin}, \&
  {Petrov}}]{Artemenko12}
{Artemenko}, S.~A., {Grankin}, K.~N., \& {Petrov}, P.~P. 2012, Astronomy
  Letters, 38, 783

\bibitem[{{Bast} {et~al.}(2011){Bast}, {Brown}, {Herczeg}, {van Dishoeck}, \&
  {Pontoppidan}}]{Bast11}
{Bast}, J.~E., {Brown}, J.~M., {Herczeg}, G.~J., {van Dishoeck}, E.~F., \&
  {Pontoppidan}, K.~M. 2011, \aap, 527, A119

\bibitem[{{Byrne}(1986)}]{byrne86}
{Byrne}, P.~B. 1986, Irish Astronomical Journal, 17, 294

\bibitem[{{Cohen} \& {Kuhi}(1979)}]{cohen79}
{Cohen}, M. \& {Kuhi}, L.~V. 1979, \apjs, 41, 743

\bibitem[{{Dai} {et~al.}(2010){Dai}, {Wilner}, {Andrews}, \& {Ohashi}}]{Dai10}
{Dai}, Y., {Wilner}, D.~J., {Andrews}, S.~M., \& {Ohashi}, N. 2010, \aj, 139,
  626

\bibitem[{{de Geus} {et~al.}(1989){de Geus}, {de Zeeuw}, \& {Lub}}]{DeGeus89}
{de Geus}, E.~J., {de Zeeuw}, P.~T., \& {Lub}, J. 1989, \aap, 216, 44

\bibitem[{{de la Reza} {et~al.}(1989){de la Reza}, {Torres}, {Quast},
  {Castilho}, \& {Vieira}}]{delareza89}
{de la Reza}, R., {Torres}, C.~A.~O., {Quast}, G., {Castilho}, B.~V., \&
  {Vieira}, G.~L. 1989, \apjl, 343, L61

\bibitem[{{Dent} {et~al.}(2005){Dent}, {Greaves}, \& {Coulson}}]{Dent05}
{Dent}, W.~R.~F., {Greaves}, J.~S., \& {Coulson}, I.~M. 2005, \mnras, 359, 663

\bibitem[{{di Folco} {et~al.}(2007){di Folco}, {Absil}, {Augereau},
  {M{\'e}rand}, {Coud{\'e} du Foresto}, {Th{\'e}venin}, {Defr{\`e}re},
  {Kervella}, {ten Brummelaar}, {McAlister}, {Ridgway}, {Sturmann}, {Sturmann},
  \& {Turner}}]{Difolco07}
{di Folco}, E., {Absil}, O., {Augereau}, J.-C., {et~al.} 2007, \aap, 475, 243

\bibitem[{{Donati} {et~al.}(2011){Donati}, {Gregory}, {Montmerle}, {Maggio},
  {Argiroffi}, {Sacco}, {Hussain}, {Kastner}, {Alencar}, {Audard}, {Bouvier},
  {Damiani}, {G{\"u}del}, {Huenemoerder}, \& {Wade}}]{Donati11}
{Donati}, J.-F., {Gregory}, S.~G., {Montmerle}, T., {et~al.} 2011, \mnras, 417,
  1747

\bibitem[{{Dullemond} \& {Monnier}(2010)}]{Dullemond10}
{Dullemond}, C.~P. \& {Monnier}, J.~D. 2010, \araa, 48, 205

\bibitem[{{Dyck} {et~al.}(1982){Dyck}, {Simon}, \& {Zuckerman}}]{dyck82}
{Dyck}, H.~M., {Simon}, T., \& {Zuckerman}, B. 1982, \apjl, 255, L103

\bibitem[{{Eisner} {et~al.}(2014){Eisner}, {Hillenbrand}, \&
  {Stone}}]{Eisner14}
{Eisner}, J.~A., {Hillenbrand}, L.~A., \& {Stone}, J.~M. 2014, mnras, 443, 1916

\bibitem[{{Eisner} {et~al.}(2005){Eisner}, {Hillenbrand}, {White}, {Akeson}, \&
  {Sargent}}]{Eisner05}
{Eisner}, J.~A., {Hillenbrand}, L.~A., {White}, R.~J., {Akeson}, R.~L., \&
  {Sargent}, A.~I. 2005, \apj, 623, 952

\bibitem[{{Eisner} {et~al.}(2007){Eisner}, {Hillenbrand}, {White}, {Bloom},
  {Akeson}, \& {Blake}}]{Eisner07}
{Eisner}, J.~A., {Hillenbrand}, L.~A., {White}, R.~J., {et~al.} 2007, \apj,
  669, 1072

\bibitem[{{Eisner} {et~al.}(2003){Eisner}, {Lane}, {Akeson}, {Hillenbrand}, \&
  {Sargent}}]{Eisner03}
{Eisner}, J.~A., {Lane}, B.~F., {Akeson}, R.~L., {Hillenbrand}, L.~A., \&
  {Sargent}, A.~I. 2003, \apj, 588, 360

\bibitem[{{Eisner} {et~al.}(2004){Eisner}, {Lane}, {Hillenbrand}, {Akeson}, \&
  {Sargent}}]{Eisner04}
{Eisner}, J.~A., {Lane}, B.~F., {Hillenbrand}, L.~A., {Akeson}, R.~L., \&
  {Sargent}, A.~I. 2004, \apj, 613, 1049

\bibitem[{{Eisner} {et~al.}(2010){Eisner}, {Monnier}, {Woillez}, {Akeson},
  {Millan-Gabet}, {Graham}, {Hillenbrand}, {Pott}, {Ragland}, \&
  {Wizinowich}}]{Eisner10}
{Eisner}, J.~A., {Monnier}, J.~D., {Woillez}, J., {et~al.} 2010, apj, 718, 774

\bibitem[{{Figueira} {et~al.}(2012){Figueira}, {Marmier}, {Bou{\'e}}, {Lovis},
  {Santos}, {Montalto}, {Udry}, {Pepe}, \& {Mayor}}]{Figueira12}
{Figueira}, P., {Marmier}, M., {Bou{\'e}}, G., {et~al.} 2012, \aap, 541, A139

\bibitem[{{Forbrich} \& {Preibisch}(2007)}]{Forbrich07}
{Forbrich}, J. \& {Preibisch}, T. 2007, \aap, 475, 959

\bibitem[{{Ghez} {et~al.}(1997){Ghez}, {White}, \& {Simon}}]{ghez97}
{Ghez}, A.~M., {White}, R.~J., \& {Simon}, M. 1997, \apj, 490, 353

\bibitem[{{Gregorio-Hetem} \& {Hetem}(2002)}]{Gregorio-Hetem02}
{Gregorio-Hetem}, J. \& {Hetem}, A. 2002, \mnras, 336, 197

\bibitem[{{Haguenauer} {et~al.}(2010){Haguenauer}, {Alonso}, {Bourget},
  {Brillant}, {Gitton}, {Guisard}, {Poupar}, {Schuhler}, {Abuter}, {Andolfato},
  {Blanchard}, {Berger}, {Cortes}, {D{\'e}rie}, {Delplancke}, {di Lieto},
  {Dupuy}, {Gilli}, {Glindemann}, {Guniat}, {Huedepohl}, {Kaufer}, {Le
  Bouquin}, {L{\'e}v{\^e}que}, {M{\'e}nardi}, {M{\'e}rand}, {Morel},
  {Percheron}, {Phan Duc}, {Pino}, {Ramirez}, {Rengaswamy}, {Richichi},
  {Rivinius}, {Sahlmann}, {Schoeller}, {Schmid}, {Stefl}, {Valdes}, {van
  Belle}, {Wehner}, \& {Wittkowski}}]{Haguenauer10}
{Haguenauer}, P., {Alonso}, J., {Bourget}, P., {et~al.} 2010, in Society of
  Photo-Optical Instrumentation Engineers (SPIE) Conference Series, Vol. 7734,
  Society of Photo-Optical Instrumentation Engineers (SPIE) Conference Series

\bibitem[{{Herbig}(1973)}]{Herbig73}
{Herbig}, G.~H. 1973, \apj, 182, 129

\bibitem[{{Herczeg} \& {Hillenbrand}(2014)}]{Herczeg14}
{Herczeg}, G.~J. \& {Hillenbrand}, L.~A. 2014, \apj, 786, 97

\bibitem[{{Hughes} {et~al.}(1994){Hughes}, {Hartigan}, {Krautter}, \&
  {Kelemen}}]{hughes94}
{Hughes}, J., {Hartigan}, P., {Krautter}, J., \& {Kelemen}, J. 1994, \aj, 108,
  1071

\bibitem[{{Kalas} {et~al.}(2008){Kalas}, {Graham}, {Chiang}, {Fitzgerald},
  {Clampin}, {Kite}, {Stapelfeldt}, {Marois}, \& {Krist}}]{Kalas2008}
{Kalas}, P., {Graham}, J.~R., {Chiang}, E., {et~al.} 2008, Science, 322, 1345

\bibitem[{{Lafrasse} {et~al.}(2010){Lafrasse}, {Mella}, {Bonneau}, {Duvert},
  {Delfosse}, {Chesneau}, \& {Chelli}}]{Lafrasse10}
{Lafrasse}, S., {Mella}, G., {Bonneau}, D., {et~al.} 2010, in Society of
  Photo-Optical Instrumentation Engineers (SPIE) Conference Series, Vol. 7734,
  Society of Photo-Optical Instrumentation Engineers (SPIE) Conference Series

\bibitem[{{Lagrange} {et~al.}(2010){Lagrange}, {Bonnefoy}, {Chauvin}, {Apai},
  {Ehrenreich}, {Boccaletti}, {Gratadour}, {Rouan}, {Mouillet}, {Lacour}, \&
  {Kasper}}]{Lagrange2010}
{Lagrange}, A.-M., {Bonnefoy}, M., {Chauvin}, G., {et~al.} 2010, Science, 329,
  57

\bibitem[{{Le Bouquin} {et~al.}(2011){Le Bouquin}, {Berger}, {Lazareff},
  {Zins}, {Haguenauer}, {Jocou}, {Kern}, {Millan-Gabet}, {Traub}, {Absil},
  {Augereau}, {Benisty}, {Blind}, {Bonfils}, {Bourget}, {Delboulbe},
  {Feautrier}, {Germain}, {Gitton}, {Gillier}, {Kiekebusch}, {Kluska},
  {Knudstrup}, {Labeye}, {Lizon}, {Monin}, {Magnard}, {Malbet}, {Maurel},
  {M{\'e}nard}, {Micallef}, {Michaud}, {Montagnier}, {Morel}, {Moulin},
  {Perraut}, {Popovic}, {Rabou}, {Rochat}, {Rojas}, {Roussel}, {Roux},
  {Stadler}, {Stefl}, {Tatulli}, \& {Ventura}}]{Lebouquin11}
{Le Bouquin}, J.-B., {Berger}, J.-P., {Lazareff}, B., {et~al.} 2011, \aap, 535,
  A67

\bibitem[{{Leinert} {et~al.}(1994){Leinert}, {Richichi}, {Weitzel}, \&
  {Haas}}]{leinert94}
{Leinert}, C., {Richichi}, A., {Weitzel}, N., \& {Haas}, M. 1994, in
  Astronomical Society of the Pacific Conference Series, Vol.~62, The Nature
  and Evolutionary Status of Herbig Ae/Be Stars, ed. P.~S. {The}, M.~R.
  {Perez}, \& E.~P.~J. {van den Heuvel}, 155

\bibitem[{{Liu} {et~al.}(2011){Liu}, {Zhang}, {Wu}, {Qin}, \& {Miller}}]{Liu11}
{Liu}, T., {Zhang}, H., {Wu}, Y., {Qin}, S.-L., \& {Miller}, M. 2011, \apj,
  734, 22

\bibitem[{{Lommen} {et~al.}(2007){Lommen}, {Wright}, {Maddison},
  {J{\o}rgensen}, {Bourke}, {van Dishoeck}, {Hughes}, {Wilner}, {Burton}, \&
  {van Langevelde}}]{Lommen07}
{Lommen}, D., {Wright}, C.~M., {Maddison}, S.~T., {et~al.} 2007, \aap, 462, 211

\bibitem[{{Luhman}(2007)}]{Luhman07}
{Luhman}, K.~L. 2007, \apjs, 173, 104

\bibitem[{{Malo} {et~al.}(2013){Malo}, {Doyon}, {Lafreni{\`e}re}, {Artigau},
  {Gagn{\'e}}, {Baron}, \& {Riedel}}]{Malo13}
{Malo}, L., {Doyon}, R., {Lafreni{\`e}re}, D., {et~al.} 2013, \apj, 762, 88

\bibitem[{{Manoj} {et~al.}(2006){Manoj}, {Bhatt}, {Maheswar}, \&
  {Muneer}}]{Manoj06}
{Manoj}, P., {Bhatt}, H.~C., {Maheswar}, G., \& {Muneer}, S. 2006, \apj, 653,
  657

\bibitem[{{Marois} {et~al.}(2010){Marois}, {Zuckerman}, {Konopacky},
  {Macintosh}, \& {Barman}}]{Marois10}
{Marois}, C., {Zuckerman}, B., {Konopacky}, Q.~M., {Macintosh}, B., \&
  {Barman}, T. 2010, \nat, 468, 1080

\bibitem[{{Mathieu} {et~al.}(1989){Mathieu}, {Walter}, \& {Myers}}]{Mathieu89}
{Mathieu}, R.~D., {Walter}, F.~M., \& {Myers}, P.~C. 1989, \aj, 98, 987

\bibitem[{{McDonald} {et~al.}(2012){McDonald}, {Zijlstra}, \&
  {Boyer}}]{Mcdonald12}
{McDonald}, I., {Zijlstra}, A.~A., \& {Boyer}, M.~L. 2012, \mnras, 427, 343

\bibitem[{{Menu} {et~al.}(2014){Menu}, {van Boekel}, {Henning}, {Chandler},
  {Linz}, {Benisty}, {Lacour}, {Min}, {Waelkens}, {Andrews}, {Calvet},
  {Carpenter}, {Corder}, {Deller}, {Greaves}, {Harris}, {Isella}, {Kwon},
  {Lazio}, {Le Bouquin}, {M{\'e}nard}, {Mundy}, {P{\'e}rez}, {Ricci},
  {Sargent}, {Storm}, {Testi}, \& {Wilner}}]{Menu14}
{Menu}, J., {van Boekel}, R., {Henning}, T., {et~al.} 2014, \aap, 564, A93

\bibitem[{{Monnier} \& {Millan-Gabet}(2002)}]{Monnier02}
{Monnier}, J.~D. \& {Millan-Gabet}, R. 2002, \apj, 579, 694

\bibitem[{{Mugrauer} \& {Neuh{\"a}user}(2005)}]{mugraueur05}
{Mugrauer}, M. \& {Neuh{\"a}user}, R. 2005, Astronomische Nachrichten, 326, 701

\bibitem[{{Pinte} {et~al.}(2008){Pinte}, {M{\'e}nard}, {Berger}, {Benisty}, \&
  {Malbet}}]{pinte08}
{Pinte}, C., {M{\'e}nard}, F., {Berger}, J.~P., {Benisty}, M., \& {Malbet}, F.
  2008, \apjl, 673, L63

\bibitem[{{Reipurth} {et~al.}(1996){Reipurth}, {Pedrosa}, \&
  {Lago}}]{Reipurth96}
{Reipurth}, B., {Pedrosa}, A., \& {Lago}, M.~T.~V.~T. 1996, \aaps, 120, 229

\bibitem[{{Romero} {et~al.}(2012){Romero}, {Schreiber}, {Cieza},
  {Rebassa-Mansergas}, {Mer{\'{\i}}n}, {Smith Castelli}, {Allen}, \&
  {Morrell}}]{Romero12}
{Romero}, G.~A., {Schreiber}, M.~R., {Cieza}, L.~A., {et~al.} 2012, \apj, 749,
  79

\bibitem[{{Sartori} {et~al.}(2003){Sartori}, {Lepine}, \& {Dias}}]{Sartori03}
{Sartori}, M.~J., {Lepine}, J.~R.~D., \& {Dias}, W.~S. 2003, VizieR Online Data
  Catalog, 340, 40913

\bibitem[{{Stempels} {et~al.}(2007){Stempels}, {Gahm}, \&
  {Petrov}}]{Stempels07}
{Stempels}, H.~C., {Gahm}, G.~F., \& {Petrov}, P.~P. 2007, \aap, 461, 253

\bibitem[{{Torres} {et~al.}(2006){Torres}, {Quast}, {da Silva}, {de La Reza},
  {Melo}, \& {Sterzik}}]{Torres06}
{Torres}, C.~A.~O., {Quast}, G.~R., {da Silva}, L., {et~al.} 2006, \aap, 460,
  695

\bibitem[{{van Leeuwen}(2007)}]{VanLeeuwen07}
{van Leeuwen}, F. 2007, \aap, 474, 653

\bibitem[{{Vural} {et~al.}(2012){Vural}, {Kreplin}, {Kraus}, {Weigelt},
  {Driebe}, {Benisty}, {Dugu{\'e}}, {Massi}, {Monin}, \& {Vannier}}]{Vural12}
{Vural}, J., {Kreplin}, A., {Kraus}, S., {et~al.} 2012, \aap, 543, A162

\bibitem[{{Wahhaj} {et~al.}(2010){Wahhaj}, {Cieza}, {Koerner}, {Stapelfeldt},
  {Padgett}, {Case}, {Keller}, {Mer{\'{\i}}n}, {Evans}, {Harvey}, {Sargent},
  {van Dishoeck}, {Allen}, {Blake}, {Brooke}, {Chapman}, {Mundy}, \&
  {Myers}}]{Wahhaj10}
{Wahhaj}, Z., {Cieza}, L., {Koerner}, D.~W., {et~al.} 2010, \apj, 724, 835

\bibitem[{{Whittet} {et~al.}(1997){Whittet}, {Prusti}, {Franco}, {Gerakines},
  {Kilkenny}, {Larson}, \& {Wesselius}}]{Whittet97}
{Whittet}, D.~C.~B., {Prusti}, T., {Franco}, G.~A.~P., {et~al.} 1997, \aap,
  327, 1194

\bibitem[{{Yang} {et~al.}(2012){Yang}, {Herczeg}, {Linsky}, {Brown},
  {Johns-Krull}, {Ingleby}, {Calvet}, {Bergin}, \& {Valenti}}]{Yang12}
{Yang}, H., {Herczeg}, G.~J., {Linsky}, J.~L., {et~al.} 2012, \apj, 744, 121

\end{thebibliography}

\appendix

\section{Observations log. }
\label{app : log}

\onecolumn

\begin{table}[ht]
\caption{ Log of the observations. The different columns display the name, date of observation, telescopes, configuration number of spectral channels, and calibrators for each target}
\centering
\begin{tabular}{lllll|lllll}
\hline\hline\noalign{\smallskip}

 Star      &  Obs. date & Config. & Nspec & Calibrators & Star       &  Obs. date & Configuration & Nspec & Calibrators \\
 
\hline\noalign{\smallskip}
 
 GQ Lup    & 2011-08-06 & A1 G1 I1 K0   &     1 & HIP 83779   & HN Lup     & 2012-07-18 & A1 G1 I1 K0   &     1 & HIP 78754   \\
           &            &               &       & HIP 77295   &            &            &               &       & HIP 76997   \\
           &            &               &       & HIP 78118   & HT Lup     & 2011-05-21 & A1 G1 I1 K0   &     1 & HIP 79355   \\
           & 2012-07-16 & A1 G1 I1 K0   &     1 & HIP 79355   &            &            &               &       & HIP 77672   \\
           &            &               &       & HIP 78754   &            & 2011-05-22 & A1 G1 I1 K0   &     1 & HIP 77672   \\
           & 2012-07-17 & A1 G1 I1 K0   &     1 & HIP 78754   &            &            &               &       & HIP 79355   \\
           &            &               &       & HIP 78014   &            & 2012-07-16 & A1 G1 I1 K0   &     1 & HIP 79355   \\
 MY Lup    & 2012-04-16 & D0 G1 H0 I1   &     1 & HIP 77962   &            &            &               &       & HIP 77672   \\
           &            &               &       & HIP 78716   &            & 2013-05-12 & A1 B2 C1 D0   &     1 & HIP 77731   \\
 RU Lup    & 2011-08-07 & A1 G1 I1 K0   &     1 & HIP 77295   &            &            &               &       & HIP 78359   \\
           &            &               &       & HD 135549   & RY Lup     & 2011-08-07 & A1 G1 I1 K0   &     1 & HIP 77964   \\
           &            &               &       & HIP 77964   &            &            &               &       & HIP 77295   \\
           & 2012-07-17 & A1 G1 I1 K0   &     1 & HIP 78754   &            & 2013-05-12 & A1 B2 C1 D0   &     1 & HIP 78456   \\
           &            &               &       & HIP 78014   &            &            &               &       & HIP 78238   \\
           & 2012-07-18 & A1 G1 I1 K0   &     1 & HIP 78754   & V1000 Sco  & 2012-07-17 & A1 G1 I1 K0   &     1 & HIP 78551   \\
           &            &               &       & HIP 78014   &            &            &               &       & HIP 79690   \\
           & 2013-05-12 & A1 B2 C1 D0   &     1 & HIP 77388   &            & 2012-08-19 & A1 G1 I1 K0   &     1 & HIP 82722   \\
           &            &               &       & HIP 77108   &            &            &               &       & HIP 80171   \\
 V1149 Sco & 2011-05-23 & A1 G1 I1 K0   &     1 & HIP 78551   & WW Cha     & 2011-02-09 & A0 G1 I1 K0   &     1 & HIP 55237   \\
           & 2012-07-18 & A1 G1 I1 K0   &     1 & HIP 78551   &            & 2012-03-05 & A1 G1 I1 K0   &     1 & HIP 54452   \\
           &            &               &       & HIP 79690   &            &            &               &       & HIP 56876   \\
 AS 205A   & 2011-05-22 & A1 G1 I1 K0   &     1 & HIP 79377   &            & 2012-07-01 & A1 G1 I1 K0   &     1 & HD 99015    \\
           &            &               &       & HIP 77338   & V4046 Sgr  & 2011-08-06 & A1 G1 I1 K0   &     1 & HIP 89922   \\
           & 2012-06-10 & A1 G1 I1 K0   &     3 & HIP 82515   &            &            &               &       & HIP 89922   \\
           & 2012-07-19 & A1 G1 I1 K0   &     1 & HIP 77338   & V2508 Oph & 2012-07-17 & A1 G1 I1 K0   &     1 & HIP 82525   \\
           &            &               &       & HIP 79377   &            & 2012-07-19 & A1 G1 I1 K0   &     1 & HIP 82722   \\
 FK Ser    & 2011-05-22 & A1 G1 I1 K0   &     1 & HIP 91530   &            &            &               &       & HIP 82384   \\
 TWA 3A    & 2011-02-09 & A0 G1 I1 K0   &     1 & HIP 53487   &            & 2012-08-19 & A1 G1 I1 K0   &     1 & HIP 82384   \\
           &            &               &       & HIP 54547   &            &            &               &       & HIP 82722   \\
 TWA 3B    & 2011-02-09 & A0 G1 I1 K0   &     1 & HIP 53487   & V2129 Oph  & 2012-04-16 & D0 G1 H0 I1   &     1 & HIP 80784   \\
           &            &               &       & HIP 54547   &            &            &               &       & HIP 77962   \\
 TWA 07    & 2011-05-22 & A1 G1 I1 K0   &     1 & HIP 51920   &            & 2012-07-19 & A1 G1 I1 K0   &     1 & HIP 79346   \\
           &            &               &       & HIP 53631   &            &            &               &       & HIP 80355   \\
           & 2012-07-18 & A1 G1 I1 K0   &     1 & HIP 51920   & V1121 Oph  & 2011-08-07 & A1 G1 I1 K0   &     1 & HIP 79884   \\
           &            &               &       & HIP 53631   &            &            &               &       & HIP 84459   \\
 TW Hya    & 2011-02-09 & A0 G1 I1 K0   &     1 & HIP 53487   & AK Sco     & 2011-08-06 & A1 G1 I1 K0   &     1 & HIP 82046   \\
           &            &               &       & HIP 54547   &            &            &               &       & HIP 83779   \\
           & 2011-05-25 & A1 G1 I1 K0   &     1 & HIP 51920   &            &            &               &       & HIP 82046   \\
           &            &               &       & HIP 53631   &            &            &               &       & HIP 82046   \\
 S Cra     & 2011-08-06 & A1 G1 I1 K0   &     1 & HIP 93470   &            & 2012-06-10 & A1 G1 I1 K0   &     3 & HIP 82515   \\
           &            &               &       & HIP 92639   &            & 2012-07-02 & A1 G1 I1 K0   &     1 & HD 152433   \\
           & 2012-07-17 & A1 G1 I1 K0   &     1 & HIP 93611   &            &            &               &       & HD 154312   \\
           & 2012-08-19 & A1 G1 I1 K0   &     1 & HIP 92858   &            &            &               &       & HD 148841   \\
           &            &               &       & HIP 93611   &            & 2013-06-05 & A1 G1 J3 K0   &     3 & HD 152884   \\
 V 709 Cra & 2012-07-18 & A1 G1 I1 K0   &     1 & HIP 92639   &            &            &               &       & HD 155736   \\
           &            &               &       & HIP 92910   &            & 2013-06-09 & A1 G1 J3 K0   &     3 & HD 152884   \\
           & 2011-08-07 & A1 G1 I1 K0   &     1 & HIP 93470   &            &            &               &       & HD 155736   \\
 V380 Ori  & 2010-12-22 & A0 G1 I1 K0   &     1 & HD 36134    &            & 2013-06-16 & D0 G1 H0 I1   &     3 & HD 149691   \\
           &            &               &       & HD 34863    &            &            &               &       & HD 152884   \\
           &            &               &       & HIP 32076   &            & 2013-07-02 & A1 B2 C1 D0   &     3 & HD 152884   \\
           & 2010-12-23 & A0 G1 I1 K0   &     7 & HD 220986   &            &            &               &       & HD 155736   \\
           &            &               &       & HD 34863    &            &            &               &       &             \\       
           &            &               &       &             &            &            &               &       &             \\      

\noalign{\smallskip}\hline
\end{tabular}
\end{table}

\section{PIONIER data }
\label{app:images}

\begin{figure}[t]
\caption{\label{app:data} uv plane, visibilities and closure phases for all the targets observed in this survey. For each target, the different colours represent different observing dates. The closure phase panels have different  ranges, as the closure phases of each target differ greatly from one another.  The baseline of the closure phases is defined at the largest baseline of the triangle. }
\includegraphics[width=17cm]{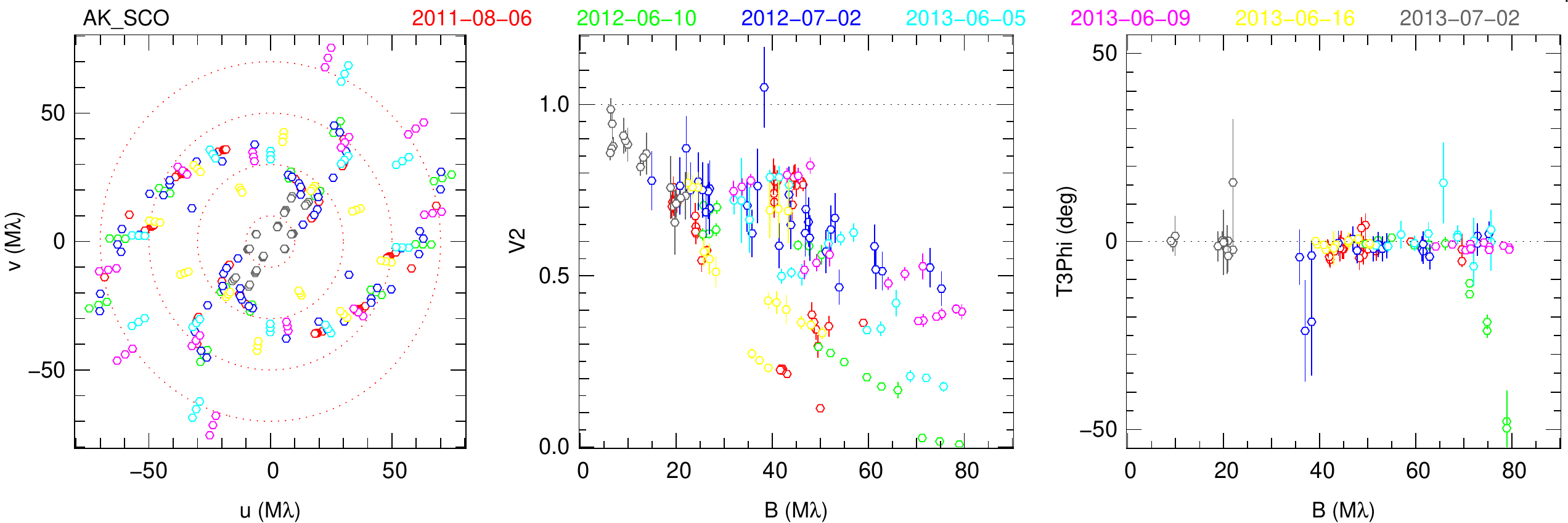}
\includegraphics[width=17cm]{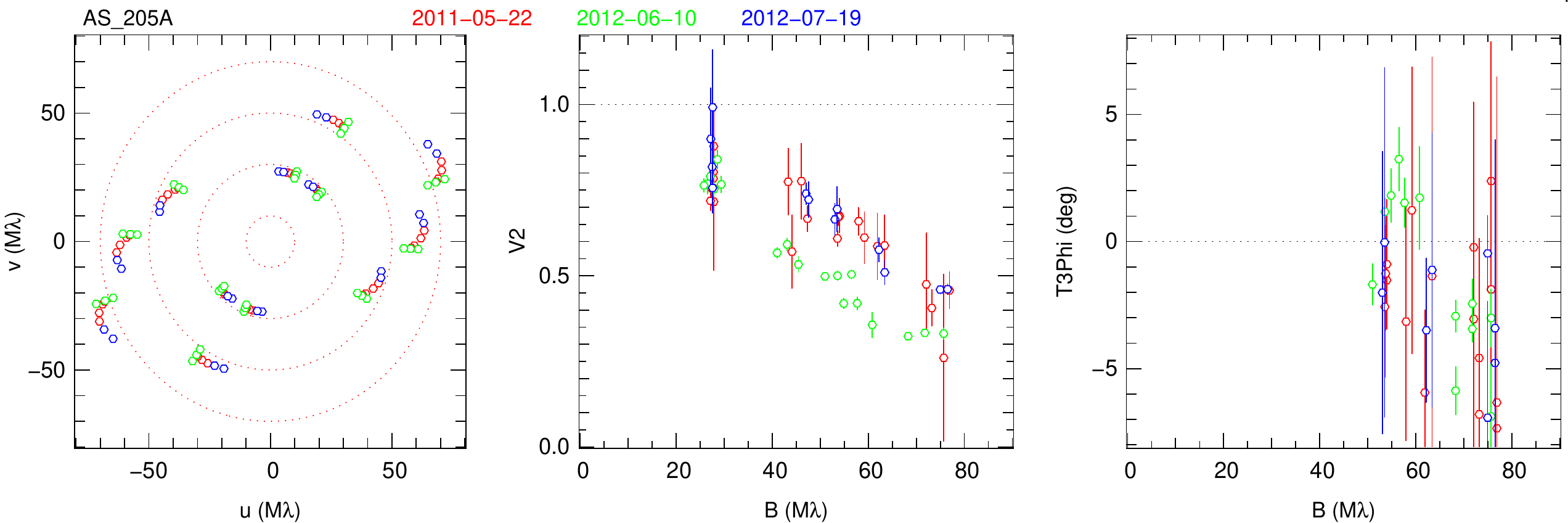}
\includegraphics[width=17cm]{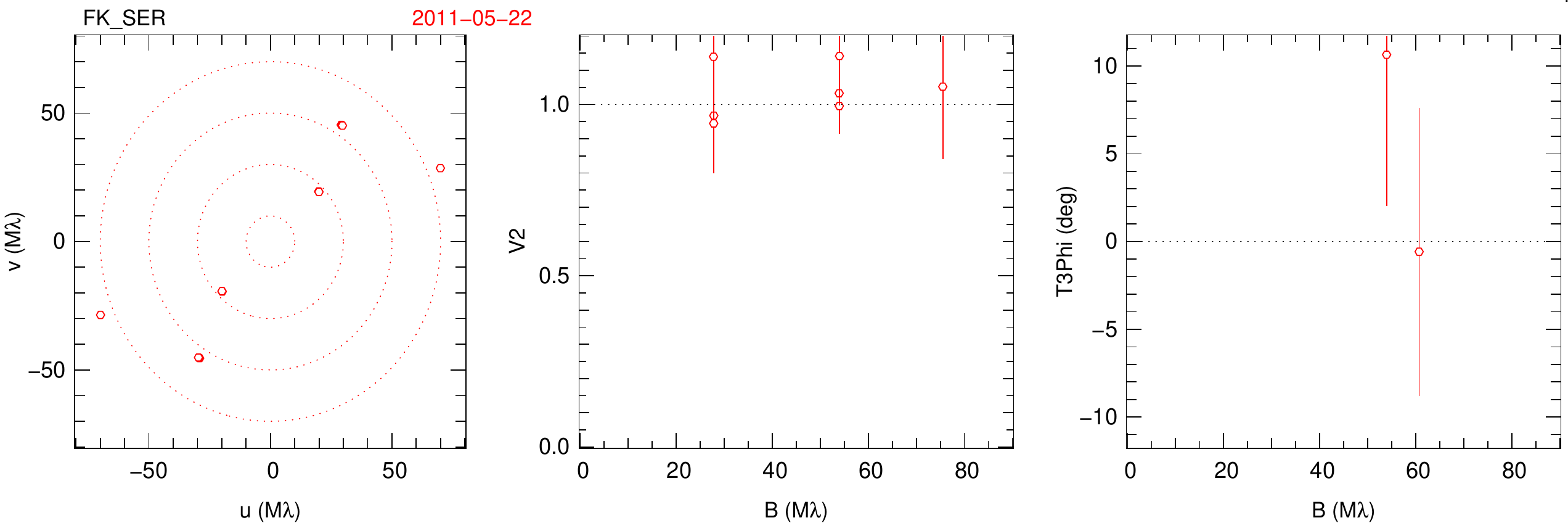}
\includegraphics[width=17cm]{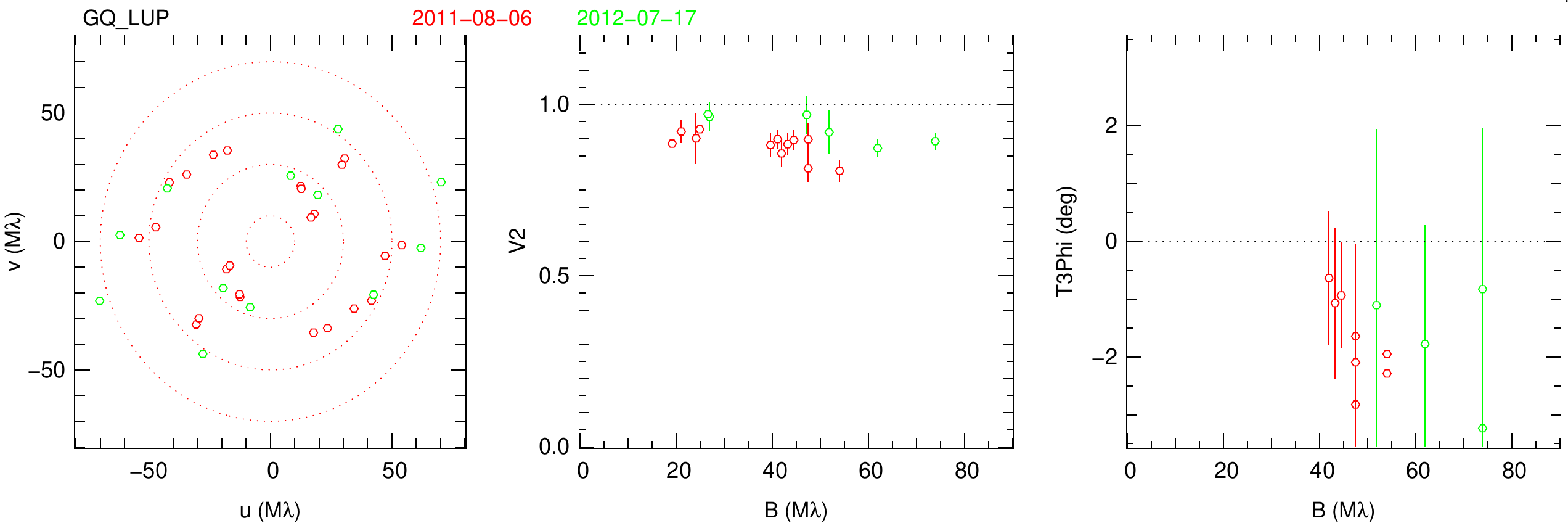}
\end{figure}
\begin{figure}[h]
\includegraphics[width=17cm]{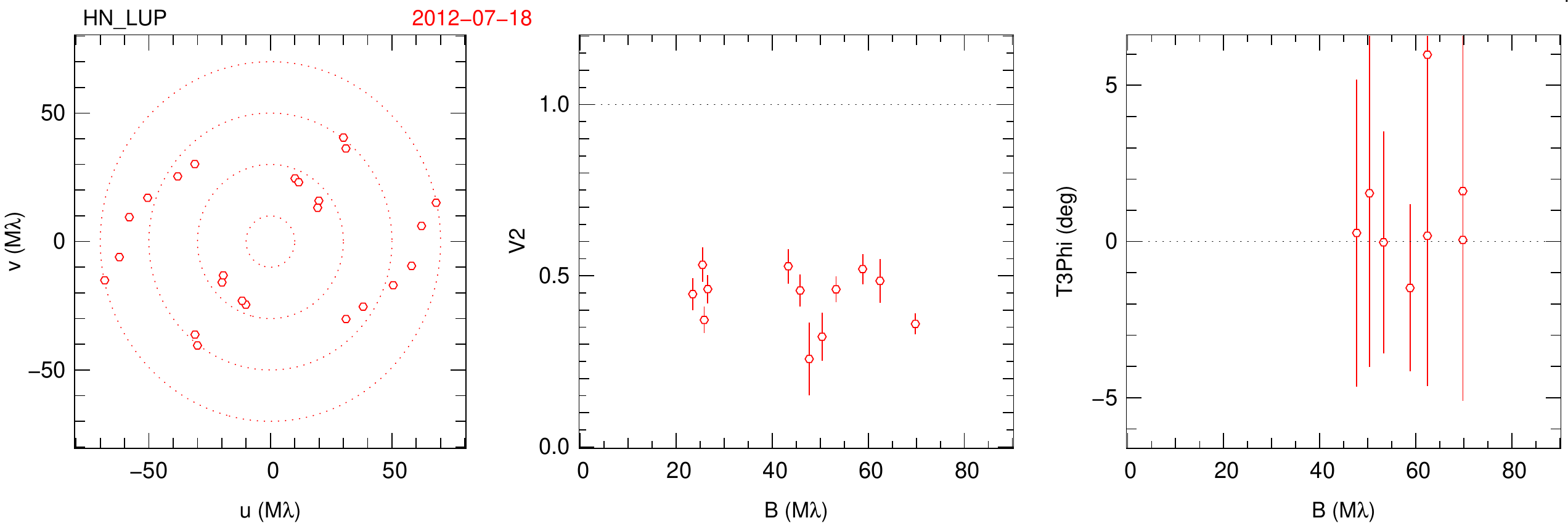}
\includegraphics[width=17cm]{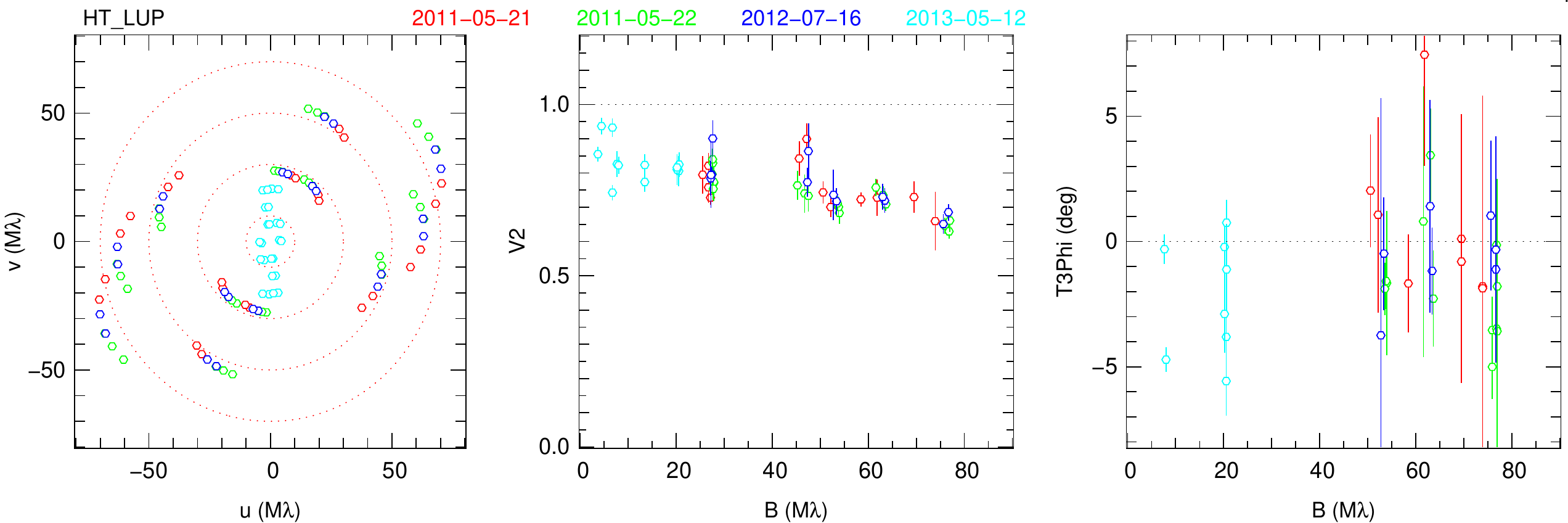}
\includegraphics[width=17cm]{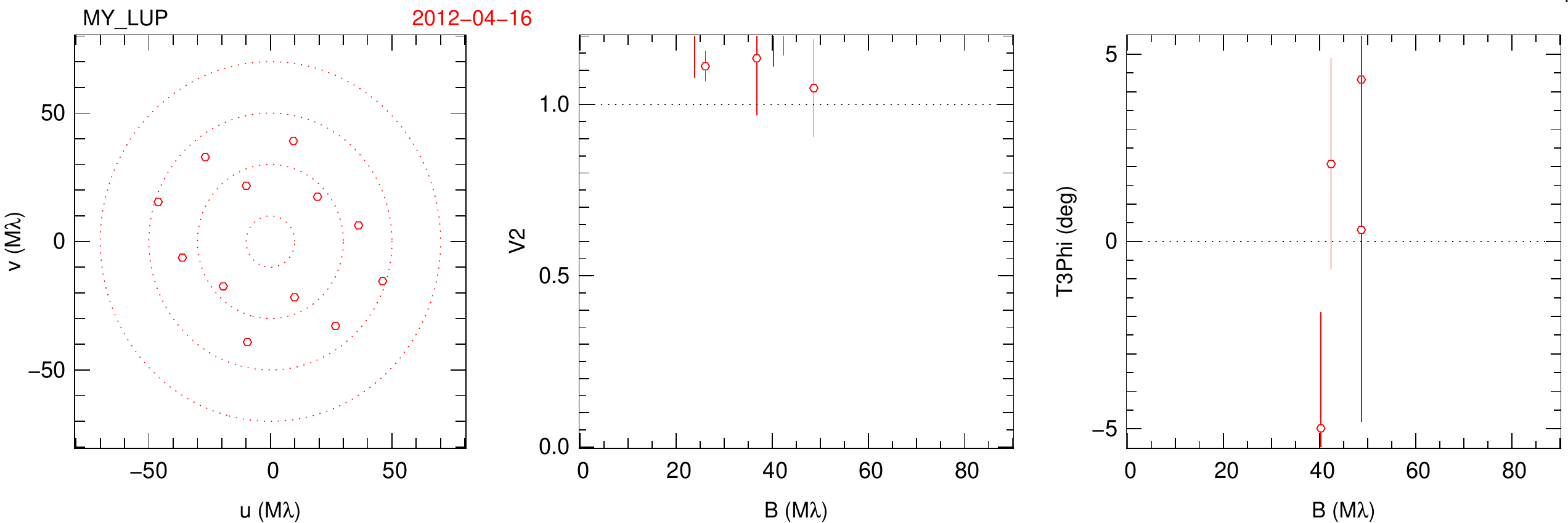}
\includegraphics[width=17cm]{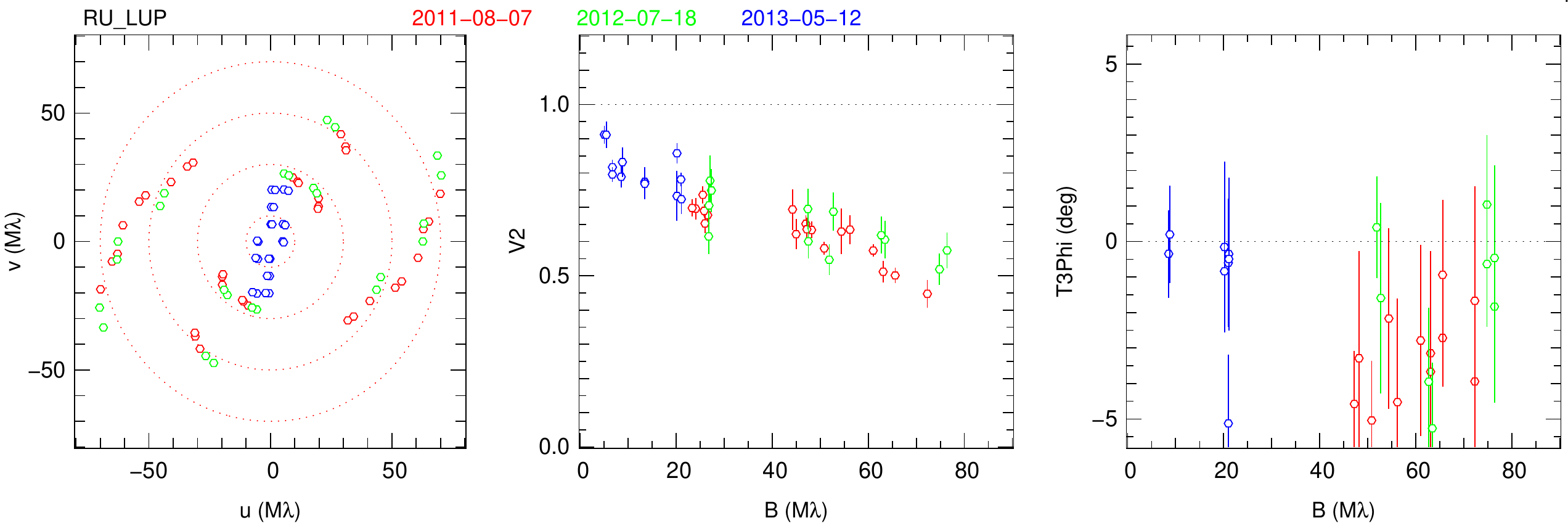}
\end{figure}
\begin{figure}[h]

\includegraphics[width=17cm]{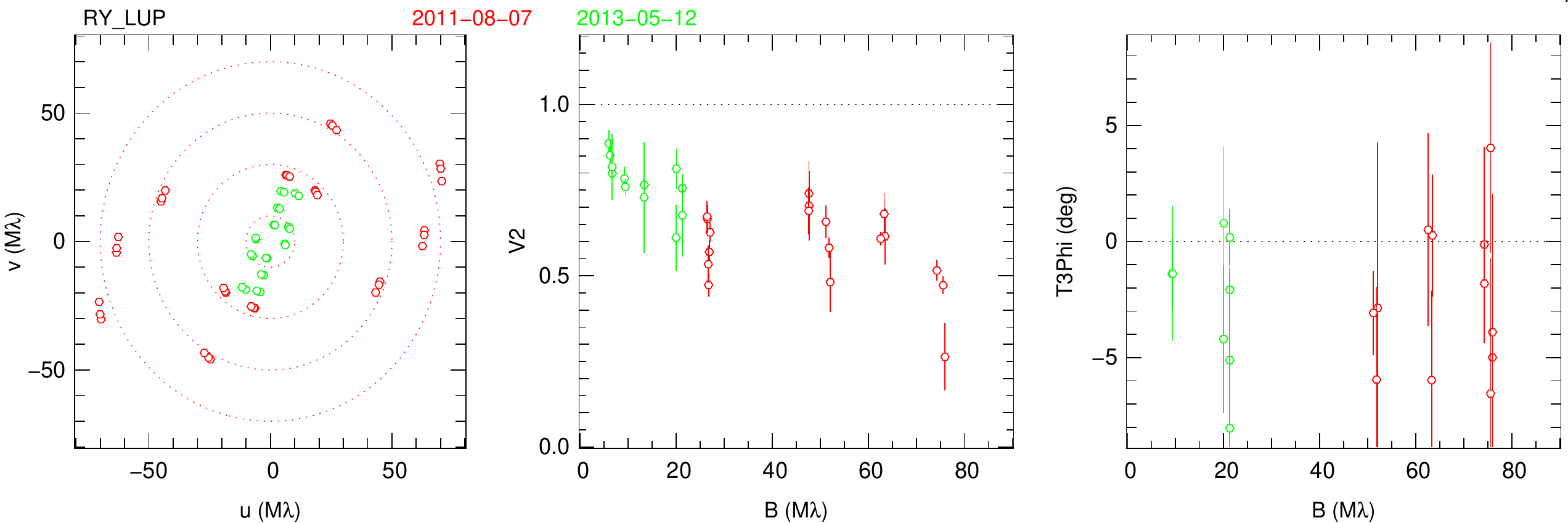}
\includegraphics[width=17cm]{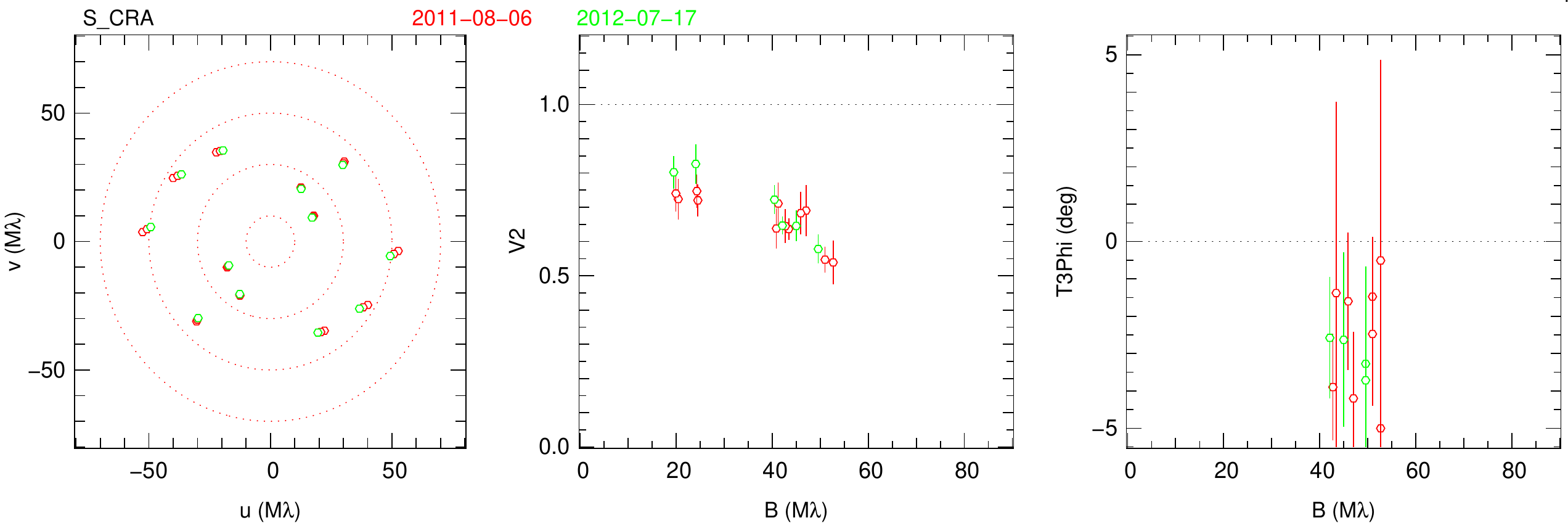}
\includegraphics[width=17cm]{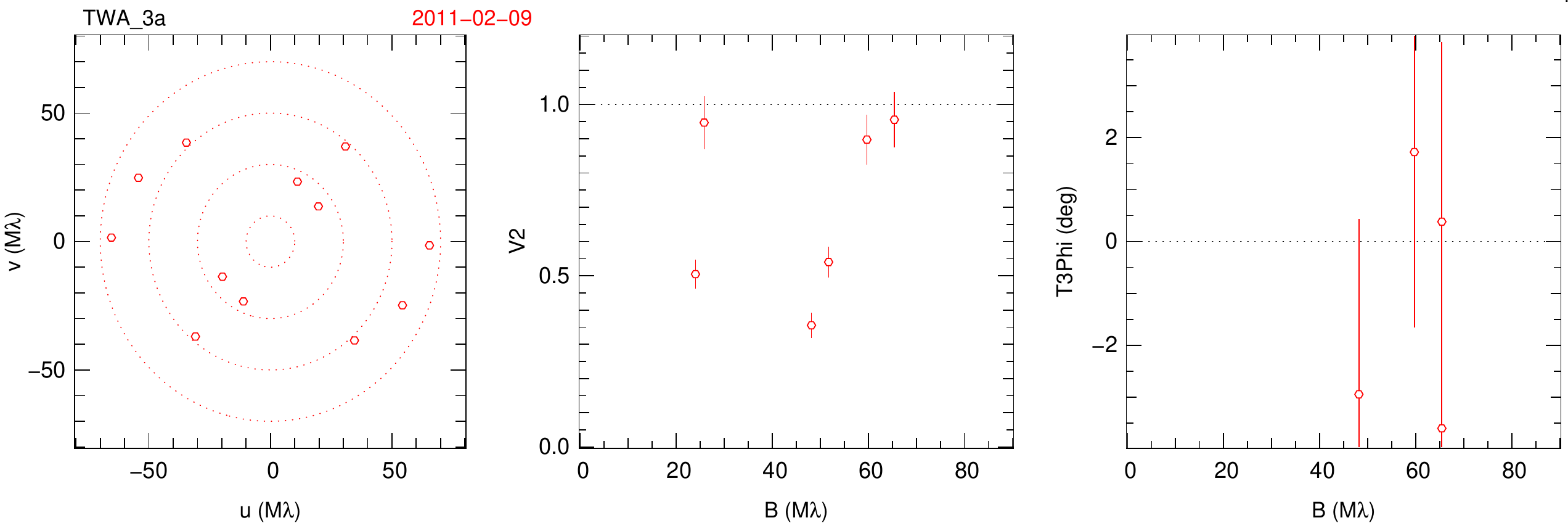}
\includegraphics[width=17cm]{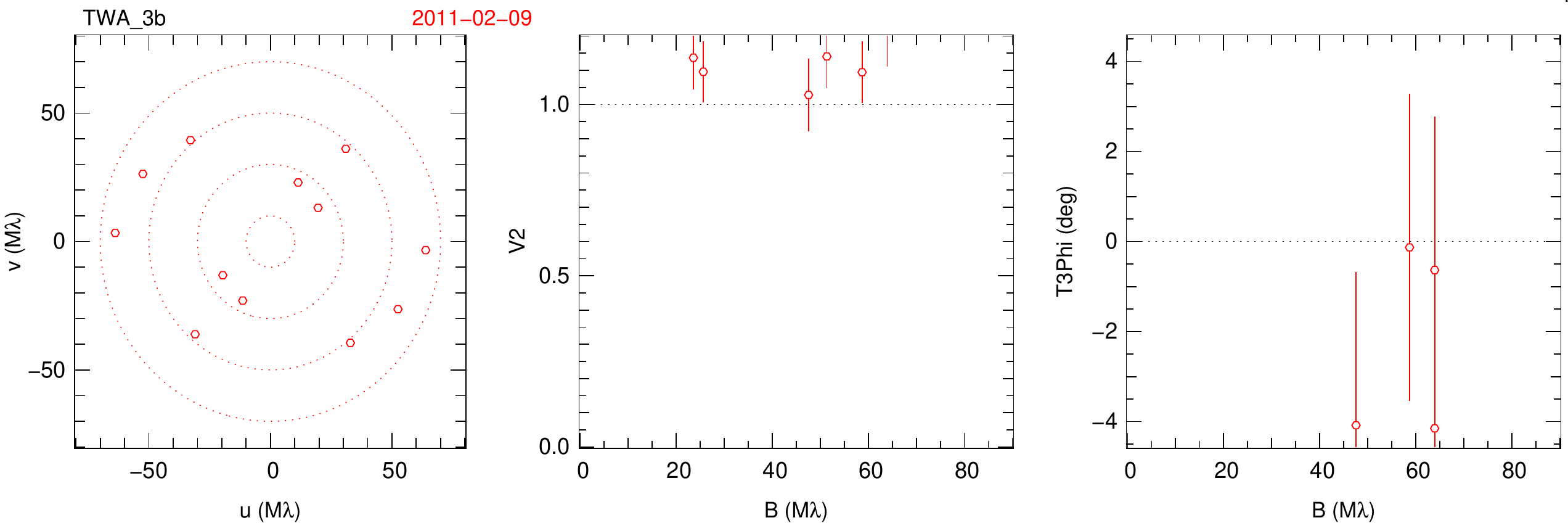}
\end{figure}
\begin{figure}[h]

\includegraphics[width=17cm]{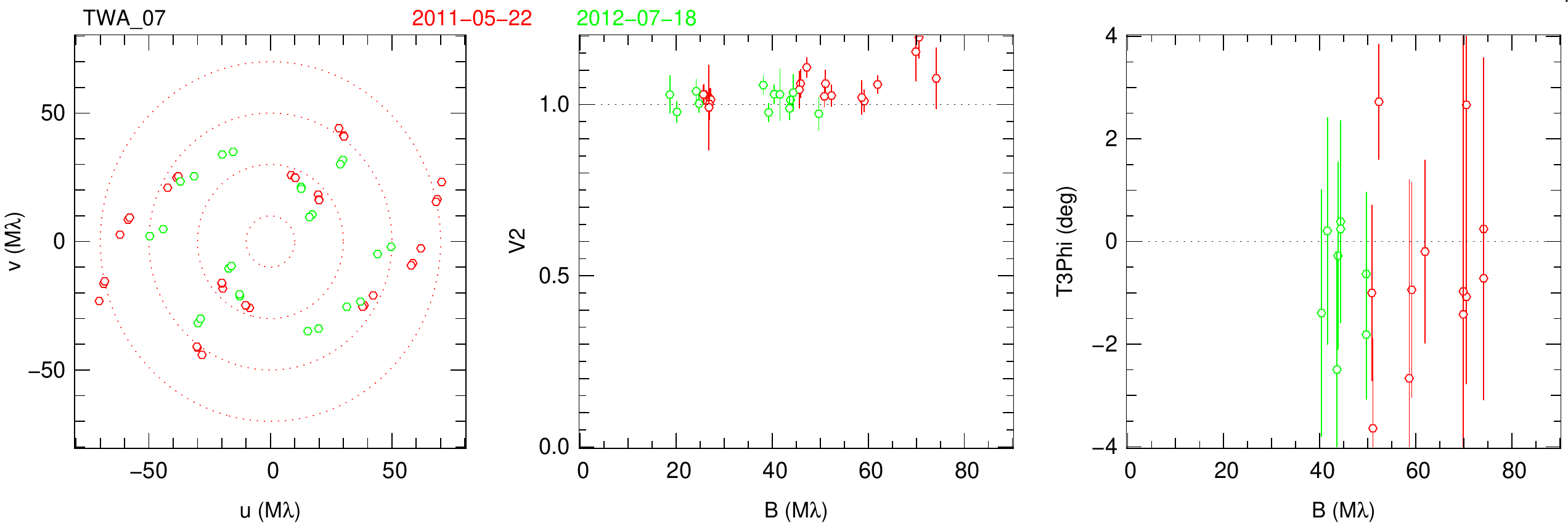}
\includegraphics[width=17cm]{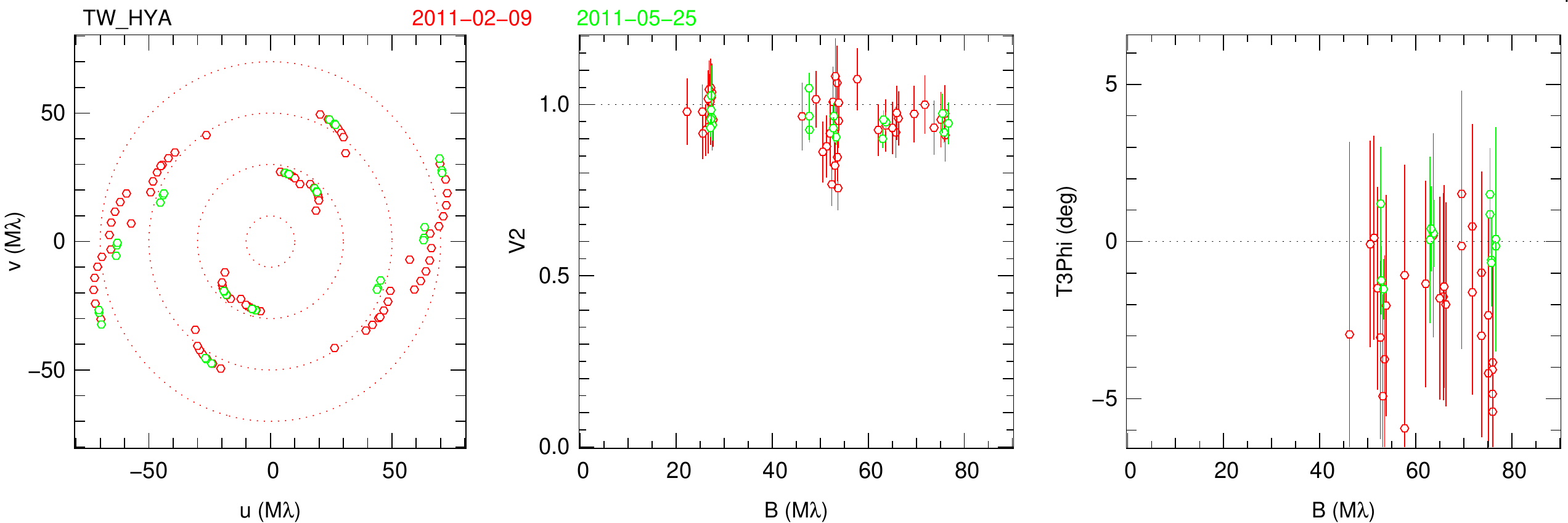}
\includegraphics[width=17cm]{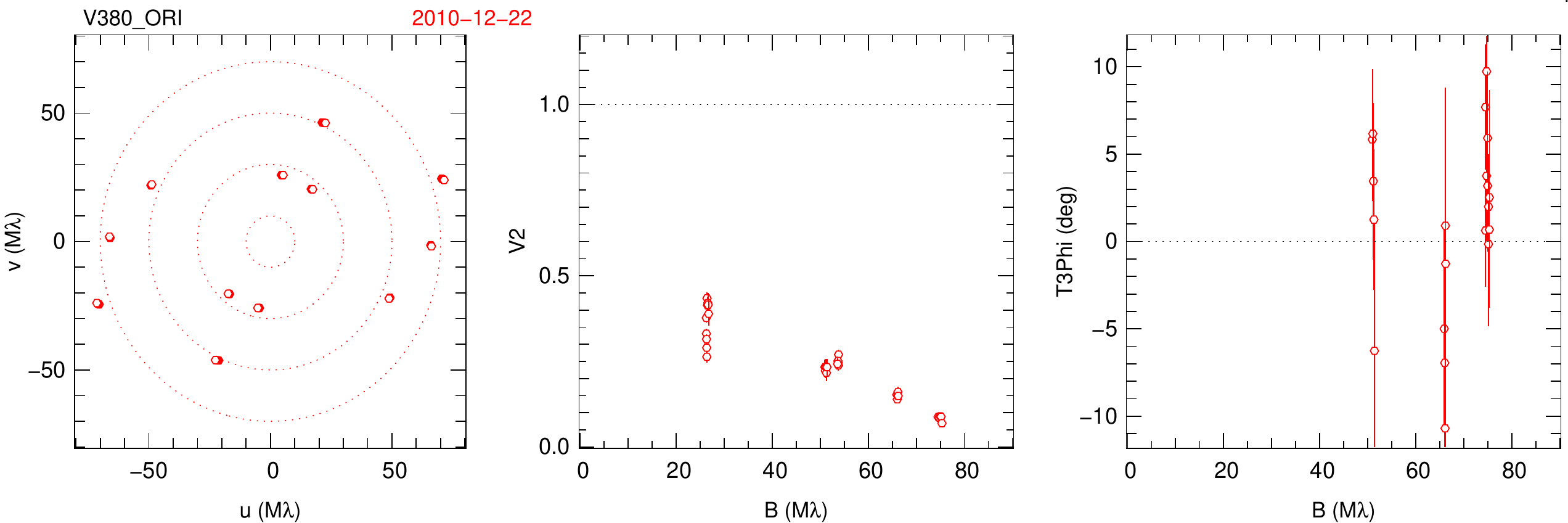}
\includegraphics[width=17cm]{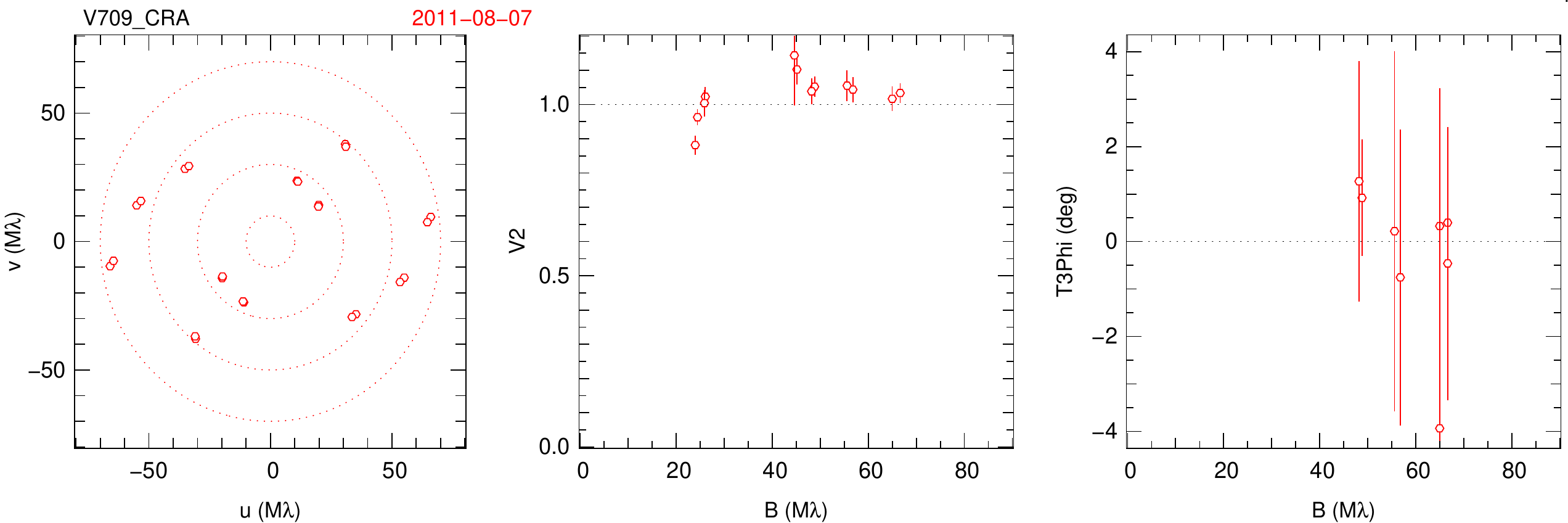}
\end{figure}
\begin{figure}[h]

\includegraphics[width=17cm]{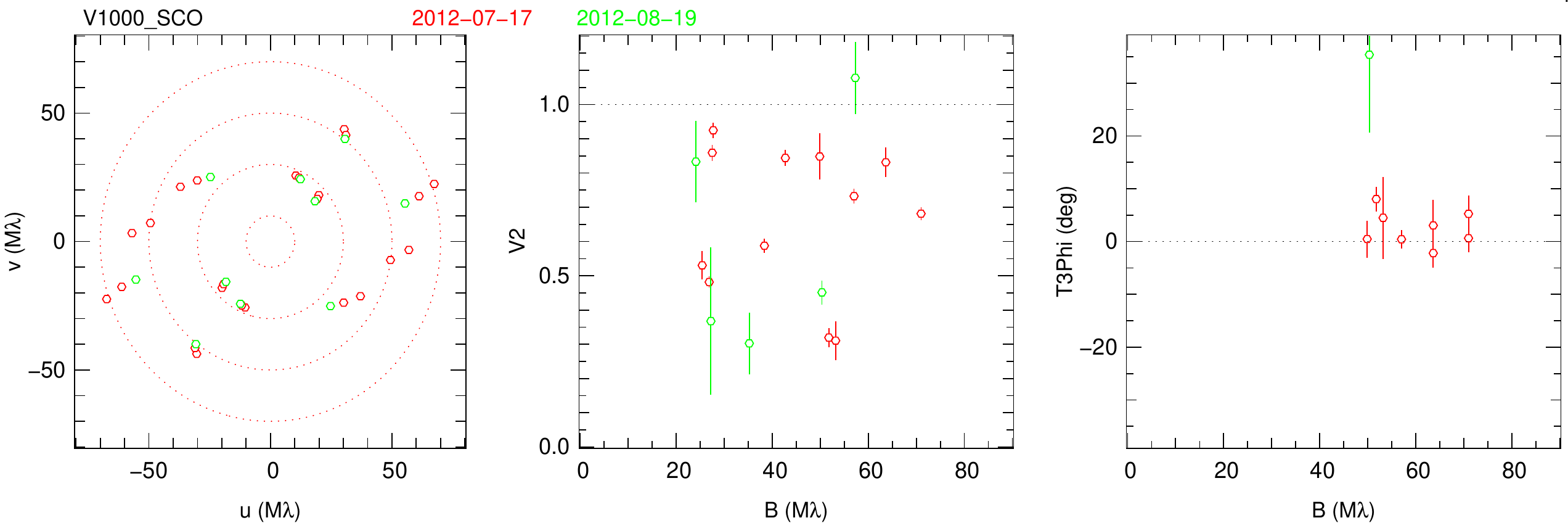}
\includegraphics[width=17cm]{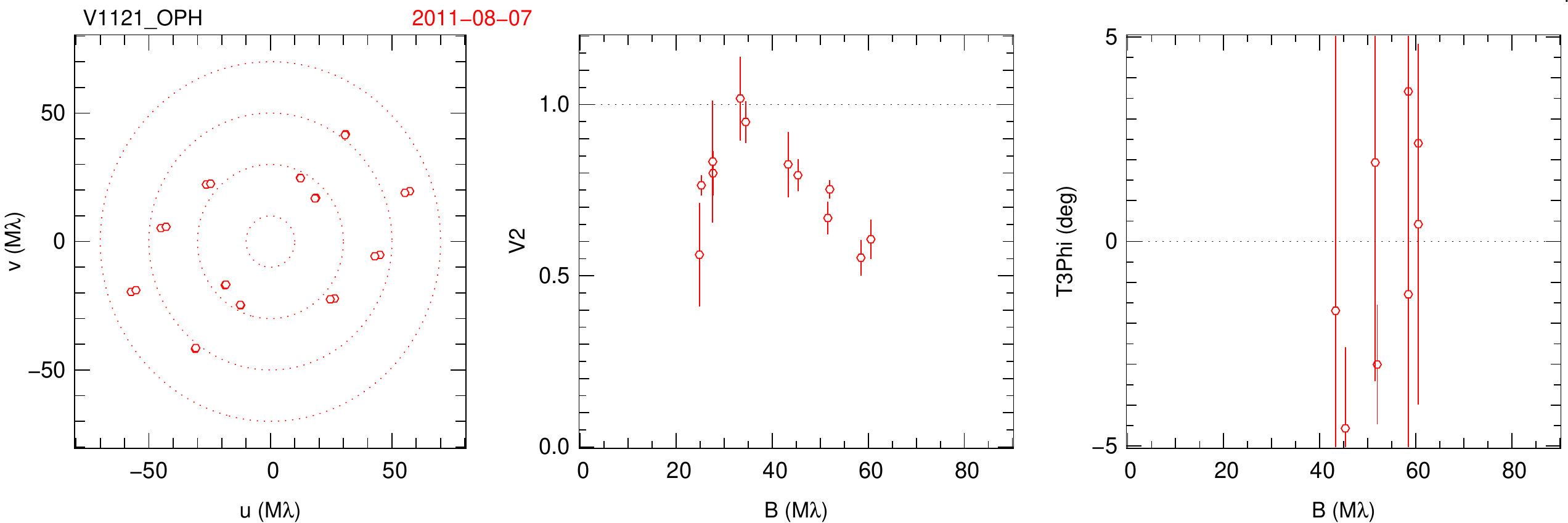}
\includegraphics[width=17cm]{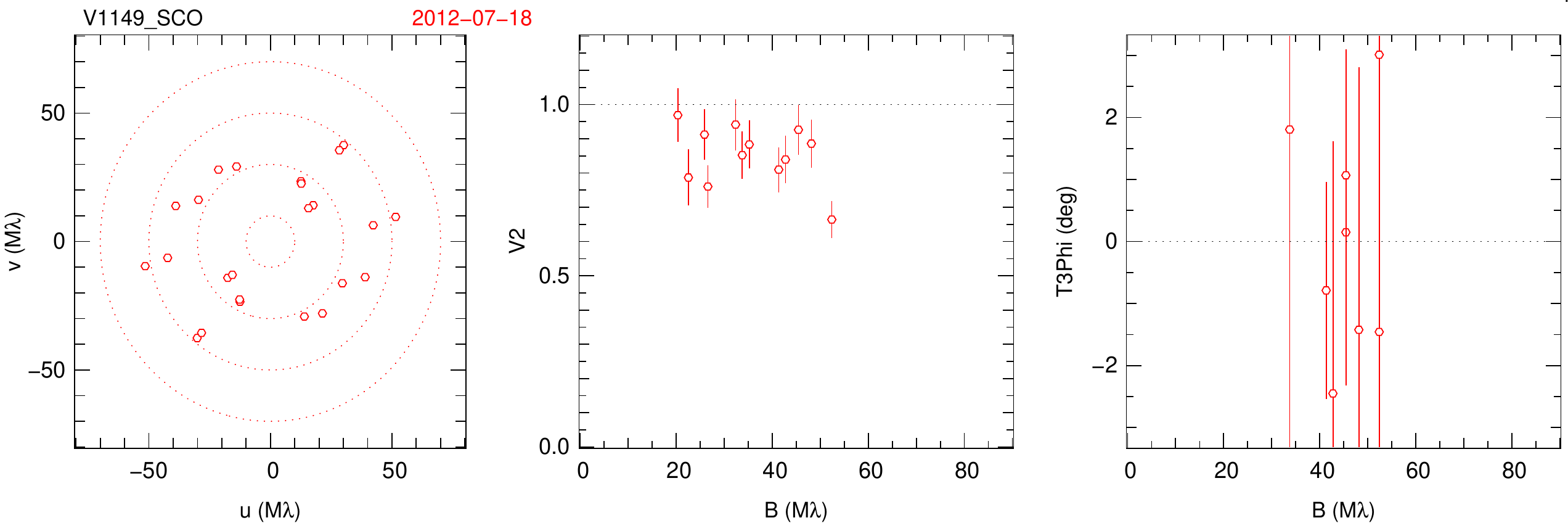}
\includegraphics[width=17cm]{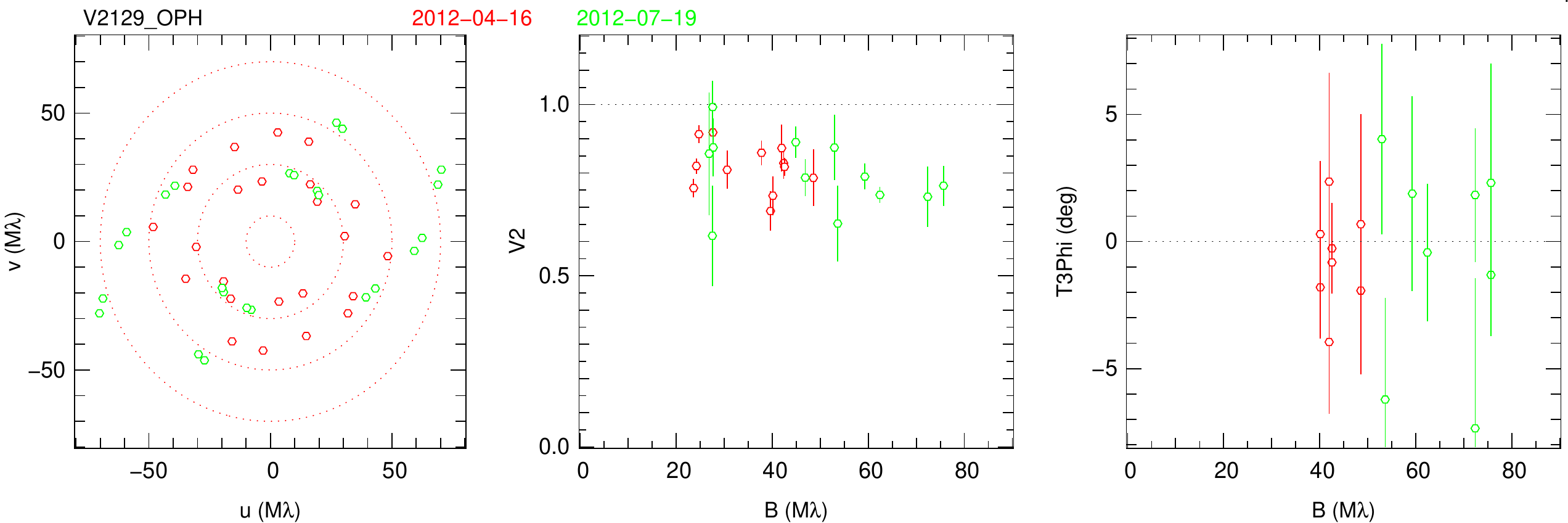}
\end{figure}
\begin{figure}[h]

\includegraphics[width=17cm]{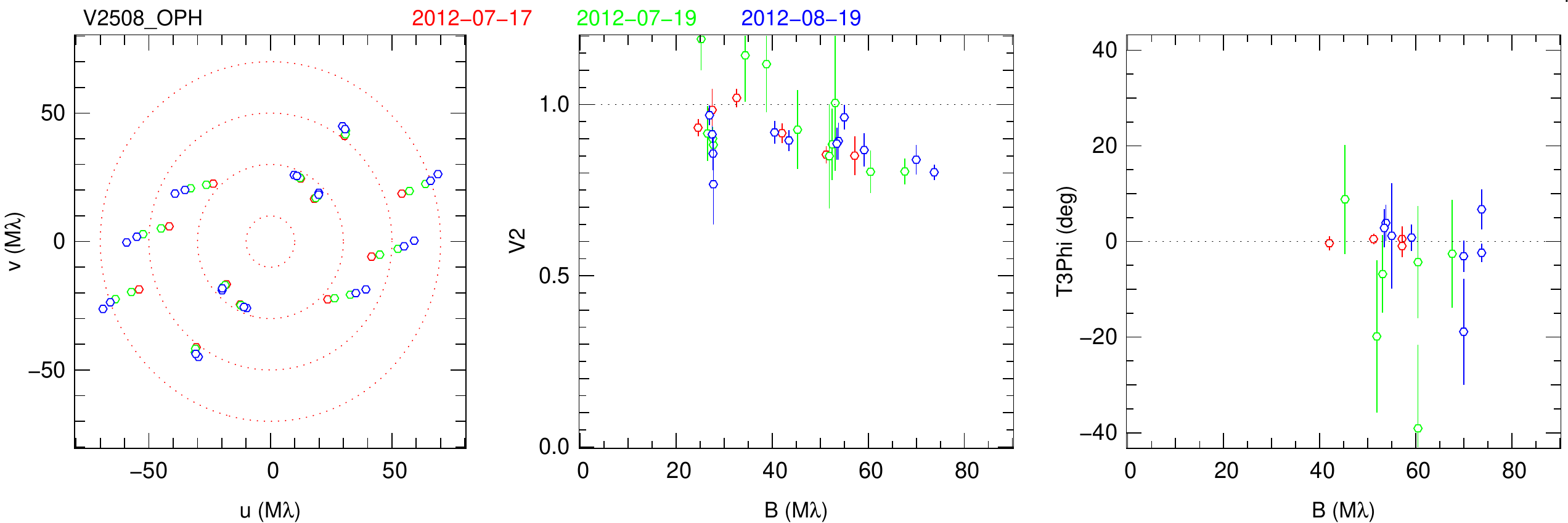}
\includegraphics[width=17cm]{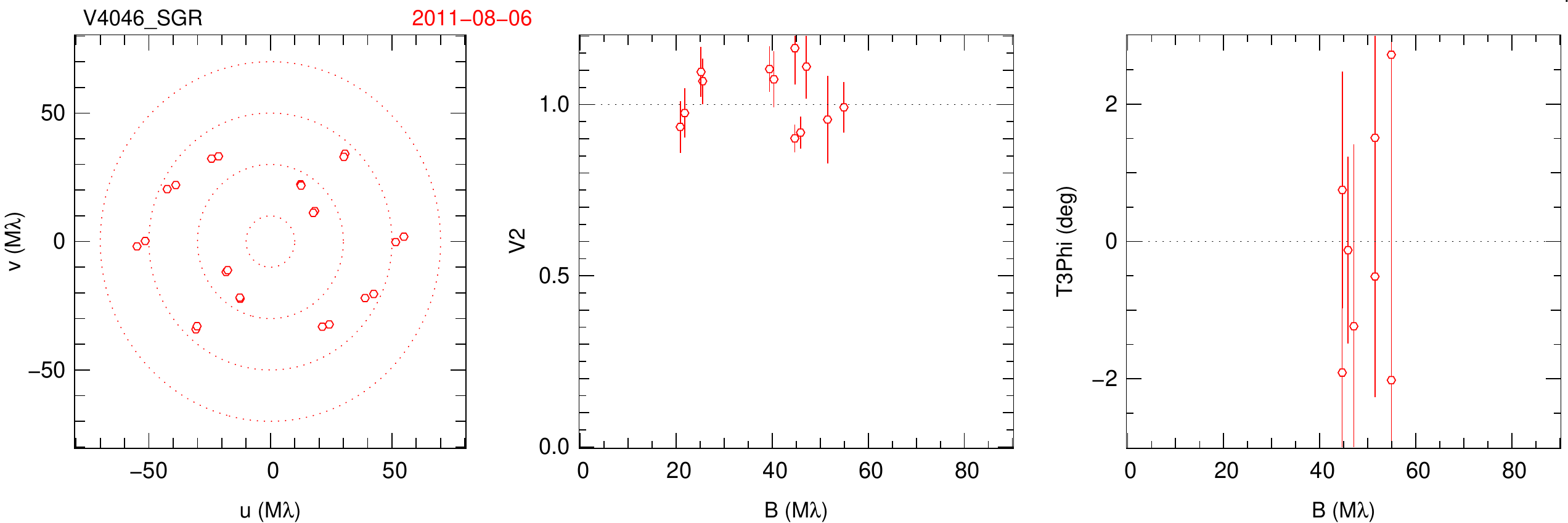}
\includegraphics[width=17cm]{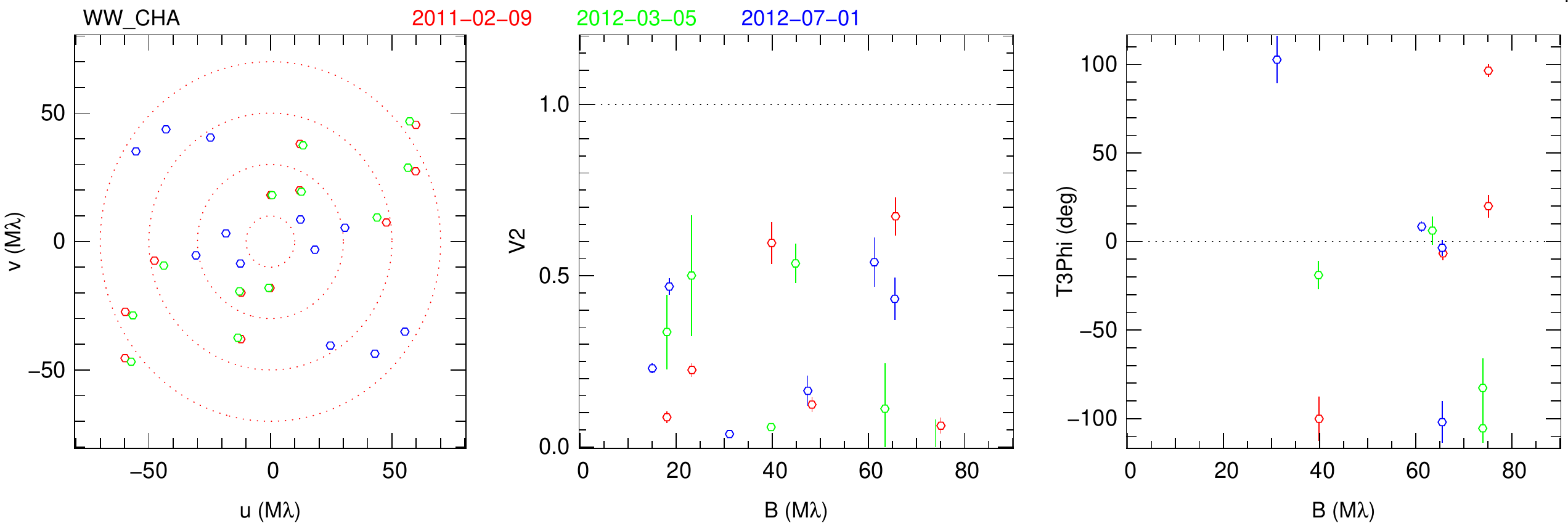}
\end{figure}

\newpage

\twocolumn

\end{document}